# Effect of molecular structure of eco-friendly glycolipid biosurfactants on the adsorption of hair-care conditioning polymers

Laura Fernández-Peña,[1] Eduardo Guzmán,[1,2,*] Fabien Leonforte,[3] Ana Serrano-Pueyo,[1]

Krzysztof Regulski,[3] Lucie Tournier-Couturier,[3] Francisco Ortega,[1,2] Ramón G. Rubio[1,2,*] and

Gustavo S. Luengo[3*]

[1]Departamento de Química Física I, Facultad de Ciencias Químicas

Universidad Complutense de Madrid, Ciudad Universitaria s/n, 28040-Madrid, Spain

[2]Instituto Pluridisciplinar, Universidad Complutense de Madrid

Paseo Juan XXIII, 1, 28040-Madrid, Spain

[3] L'Oréal Research and Innovation, Aulnay-Sous Bois, France

*To whom correspondence should be sent at: eduardogs@quim.ucm.es, rgrubio@quim.ucm.es

and gluengo@rd.loreal.com

**Abstract**

Pseudo-binary mixtures of different glycolipids, four different rhamnolipids (RL) and an alkyl polyglucoside (APG), with poly(diallyl-dimethylammonium chloride) (PDADMAC) have been studied in relation to their adsorption onto negatively charged surfaces to shed light on the impact of the molecular structure of biodegradable surfactants from natural sources (instead of synthetic surfactant, such as sodium laureth sulfate) on the adsorption of hair-care formulations. For this purpose, the self-assembly of such mixtures in aqueous solution and their adsorption onto negatively charged surfaces mimicking the negative charge of damaged hair fibres have been studied using experiments and self-consistent field (SCF) calculations. The results show that the specific physico-chemical properties of the surfactants (charge, number of sugar rings present in surfactant structure and length of the hydrocarbon length) play a main role in the control of the adsorption process, with the adsorption efficiency and hydration being improved in relation to conventional sulfate-based systems for mixtures of PDADMAC and glycolipids with the shortest alkyl chains. SCF calculations and Energy Dispersive X-Ray Spectroscopy (EDS) analysis on real hair confirmed such observations. The results allow one to assume that the characteristic of the studied surfactants, especially rhamnolipids, conditions positively the adsorption potential of polyelectrolytes in our model systems. This study provides important insights on the mechanisms underlying the performance of more complex but natural and eco-friendly washing formulations.

**Keywords**

## 1. Introduction

Surfactants are chemical compounds with application (detergency, emulsification, foaming, wetting) in several industries which ranges from the enhanced oil recovery to the food industry, and from the fabrication of drug delivery devices to the cosmetic industry [1]. However, many of the most commonly used surfactants are considered harmful for human health and the environment, which makes mandatory seeking for alternatives to replace such surfactants without compromising the properties of the final products. This is especially important considering the implementation of new legislations, e.g. European Union Regulation 2018/35 (REACH) [2], which restricts the number of chemical compounds available for industrial and consumer products. It is recommended, when it is possible, the fabrication of eco-sustainable products using components of natural origin, preferably vegetal or bacterial sources. This requires a careful examination of the physico-chemical properties and structures of the new ingredients to understand their impact in the properties of the product.

The above problem is especially important for cosmetic industry, especially when formulations for hair washing and conditioning (shampoos/conditioners) are considered. These formulations should be efficient in washing, including foaming, and repairing the hair fibers, as well as providing them with good sensorial properties, such as softness, glossiness, etc. This can be only accomplished through complex mixtures including several components, among them polyelectrolytes and surfactants bearing opposite charges [3-7]. However, some of the components are included among those which should be replaced for fulfilling the requirements of the new legislations [2], making necessary the use of new ingredients, which allow one to make formulations with similar performance than the current one, fulfilling additional requirements, such as eco-sustainable and reduced irritant character for skin and mucosa. This has moved the interest of the cosmetic industry toward the use of greener compounds, the glycolipids being example of them.

Glycolipids are biosurfactants that fulfill the eco-sustainability requirements, their biodegradability and their origin from natural sources being accounted among the main advantages of their use for substituting synthetic surfactants [8-13]. Furthermore, the good physico-chemical properties of glycolipids, as emulsifying, dispersing, foaming, wetting and coating agents, make them promising substitutes of synthetic surfactants in shampoo and conditioner formulations [10, 14, 15]. However, the generalization of the use of biosurfactants in hair care and conditioning products requires deepening in the impact of the use of these surfactants in the properties of the final produc [8, 16-18]. Therefore, the understanding of the differences in the behavior between formulations containing conventional surfactants, such as sodium laureth sulfate (SLES), and those containing biosurfactants, such as rhamnolipids, plays an important role for developing new cleansing cosmetic formulations [4, 8, 17-20]. It is worth noting that shampoos contain together with SLES, other surfactants such as the amphotheric surfactants coco-amidopropyl-betaine (CAPB) and coco-betaine (CB). However, the replacement of such surfactants is not a critical aspect for the new legislation. Therefore, this study, for sake of simplicity, will be only focused in the behavior of the effect of replacement of SLES in pseudo-binary polyelectrolyte-surfactant mixtures.

Among glycolipids, those belonging to the rhamnolipid family are probably one of the most promising options due to their chemical and structural richness. Rhamnolipids (RL) are biosurfactants with a hydrophobic moiety formed by a 3-(hydroxyalkanoyloxy)alkanoic acid bound through a glycosidic link to a hydrophilic head formed by different number of rhamnose units. Their particular structure being dependent on the bacterial strain and carbon source [21].

Adsorption of PDADMAC+RL mixtures

This study is aimed to shed light on the impact of the use of RL instead of the anionic surfactant SLES in model washing solutions formed by pseudo-binary polyelectrolyte-surfactant mixtures, paying attention to the effect of the introduction of RL in the deposition of polyelectrolyte-surfactant complexes onto solid model surfaces. More specifically, the used surfaces present similar charge density and contact angle than damaged hair fibers. This might help to understand the impact of RL in the conditioning performance of cosmetic formulations, especially because it is commonly accepted that the conditioning process relies in the deposition of different species, among them polyelectrolyte-surfactant complexes, onto the surface of damaged hair fibers [6, 22-24]. Fibers after weathering or bleaching treatments, present sulfonate groups, $SO_3^-$, on their surface reaching a density of about 2.2 $SO_3^-/nm^2$ [25]. However, the extraordinary chemical and structural complexity of real hair fibers (heterogeneity, high roughness and surface area) makes it difficult the study of the deposition processes directly onto their surface [26, 27]. The use of negatively charged flat surfaces, gold surfaces modified with negatively charged ω- mercapto-alkyl-sulfonic acid or silicon wafers enriched in silanol groups, has been found to mimic the key characteristics of hair fibers for understanding the most fundamental physico-chemical bases underlying the deposition process of conditioning agent onto hair fibers [6, 22, 28-32].

In this work, we have been carried out a systematic study of the adsorption onto negatively charged surfaces of binary PDADMAC–RL mixtures, with PDADMAC being the poly(diallyl-dimethylammonium chloride). The deposition onto negatively charged solid surfaces of RL, when they are mixed with a polycation, with respect to the performance of mixtures containing SLES has been studied. Furthermore, we have characterized the self-assembly process of polyelectrolyte-surfactant complexes in the bulk. For this purpose, the combination of experimental tools and self-consistent field (SCF) calculations has been used for understanding the effect of the molecular structure of different glycolipids. In particular, the behavior of four different RL with different hydrophobicity has been explored: a set of two RL (mono-RL and di-RL) obtained from *Pseudomonas aeruginosa* fed by glycerol leading to surfactants with alkyl tails formed 10 carbons, and a second set of two RL (mono-RL and di-RL) produced by a separate strain of *Burkholderia thailandensis* with yield in surfactant with hydrophobic chains of 14 carbons. In both cases, surfactants differing in the number of rhamnose units in the hydrophilic head were obtained: one in mono-RL, and two for di-RL. Furthermore, the replacement of RL by an alkyl poly-glucoside (APG) has also been analyzed due to the structural similarity between such surfactant and RL.

## 2. Experimental Section

**2.1 Chemicals.** Poly(diallyldimethyl)ammonium chloride (PDADMAC) was purchased from Sigma-Aldrich (Germany) with an average molecular weight in the 100-200 kDa range, and was used without further purification.

Four different RLs were extracted from *Pseudomonas aeruginosa* and *Burkholderia thailandensis* [33-35]. RL containing one or two rhamnose rings as hydrophilic heads have been chosen, mono-RL and di-RL. Among each family of RLs, surfactants with two different lengths of hydrophobic hydrocarbon chains were studied: 10 and 14 carbons. RLs are classified on the basis of the number of rhamnose rings and the length of the hydrocarbon chain as follows: mono-RL($C_{10}$), mono-RL($C_{14}$), di-RL($C_{10}$) and di-RL($C_{14}$). It is worth mentioning that RL were obtained through biotechnological methods, and conveniently isolated and purified before their use, yielding at the end of the process in pure RL samples (further details in Section S.1 of Supporting Information (SI)).

Adsorption of PDADMAC+RL mixtures

Alkyl polyglucoside (APG) and Sodium Laureth Sulfate (SLES) were purchased from Clariant (Switzerland). Scheme 1 shows the chemical structure of polymer and the different surfactants used.

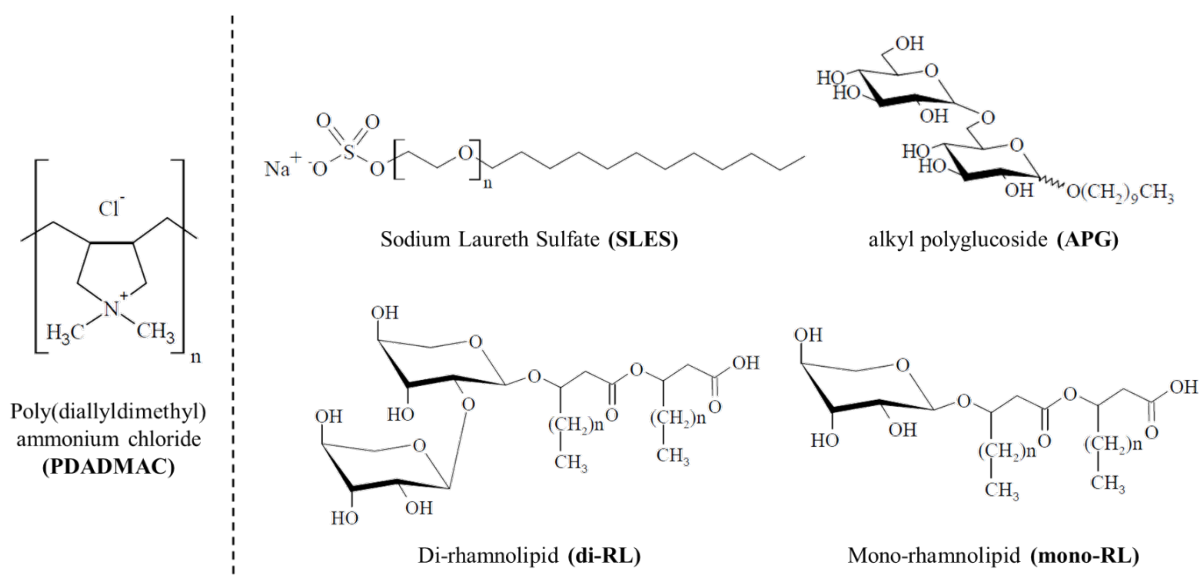

Scheme 1. Chemical structures for the polymer and surfactants used in this study. Notice that RLs were used with two different lengths of hydrocarbon chain (10 and 14). In the case of SLES n = 5.

The pH of all solutions was adjusted to 5.6, using glacial acetic acid (purity > 99 %) and the ionic strength was kept constant by adding a KCl concentration of 40 mM to match the conditions of most hair care formulations.[3] Sample preparation was carried out following a protocol adapted from the described by Llamas et al. [24] (see Section S.2, SI). It is worth mentioning that the pH, and PDADMAC (0.5 wt%) and KCl (40 mM) concentrations were chosen to match the conditions of most hair care formulations to decrease the bleaching of hair surface. Furthermore, these conditions allow minimizing the irritation of skin and mucosa [3]. Ultrapure deionized water was used for cleaning and solubilization presenting a resistivity

Adsorption of PDADMAC+RL mixtures

higher than 18 MΩ·cm and a total organic content lower than 6 ppm (Younglin 370 Series, South Korea). All experiments were carried out at 25.0 ± 0.1°C.

## 2.2 Techniques.

*a. Bulk Characterization.* Dynamic Light Scattering measurements were performed with a He-Ne laser ($\lambda = 632$ nm) in quasi-backscattering configuration ($\theta = 173°$) using a Zetasizer Nano ZS (Malvern Instrument, Ltd., United Kingdom) for evaluating the apparent diffusion coefficient, $D_{app}$, of the aggregates.

The turbidity of polyelectrolyte-surfactant mixtures was obtained from transmittance (T) measurements at 400 nm using a UV/visible spectrophotometer (HPUV 8452). The turbidity was represented as $OD_{400} = 100$-T (%).

*b. Adsorption at the liquid/solid interface.* A dissipative quartz-crystal microbalance (QCM-D) from KSV (Model QCM Z-500, Finland) fitted with quartz crystals modified by-self-assembly of a sodium salt of 3-mercapto-1-propane-sulfonic acid monolayer to obtain a negative charge surface was used. The fundamental mode ($f_0=5$ *MHz*) and odd overtones up to the 11[th] (central frequency $f_{11}=55$ *MHz*) were measured, and the acoustic thickness of the layer, $h_{ac}$, was obtained analyzing the results in terms of the model described by Voinova et al.[36].

An imaging null-ellipsometer from Nanofilm (Model EP[3], Germany) was used to determine the optical thickness of the layers, $h_{op}$. The experiments were performed using a solid/liquid cell at a fixed angle of 60°, and charged $SiO_2$ substrates (Siltronix, France) were used. The charge of $SiO_2$ substrates was obtained by treatment with *piranha* solution (30 minutes). Ellipsometric angles △ and Ψ allow the thickness of the adsorbed layers, $h_{op}$ to be calculated solving the Fresnel equation on the bases of a four slab model [37-39].

Adsorption of PDADMAC+RL mixtures

AFM measurements of dry layers deposited onto $SiO_2$ substrates were carried out in air at room temperature using a NT-MDT Ntegra Spectra (NT-MDT, Russia) in the tapping mode using a silicon tip, model RTESP (Veeco Instrument Inc, USA). It is worth mentioning that even some changes in the morphology could be expected due to the drying process, the general aspects obtained from the analysis of wet and dry samples should not change significantly the conclusions extracted from images [24].

It is worth mentioning that either thiol decorated gold surfaces and silicon wafers can be reused for team experiments, which requires a careful handling of the substrate during experiments and conservation. For the reutilization of the substrate, it is needed to perform a strict cleaning protocol involving the soaking of the substrates in ethanol in an ultrasound bath, followed by rinsing with abundant water before to immersion in *piranha* solution during 30 minutes.

Further experimental details are given in SI (Section S.2).

*C. Adsorption studies using hair fibres as substrate.* A bunch of standardized virgin/untreated hair of Caucasian origin (length 27 cm, weight 1.0 g) supplied by International Hair Importers & Products, Inc. (U.S.A.) was used for evaluating the adsorption of polymer-surfactant mixtures onto a real cosmetic substrate. Fibres were pre-treated before its use following the methodology described in SI (Section S.3.B). A Scanning Electron Microscope JEOL (model JSM 7600-F, Japan) allows imaging the surface of bleached hair fibres before and after exposure to polymer-surfactant solutions. The compositional analysis of the hair fibres was performed using an energy-dispersive X-ray spectrometer (EDS) coupled to the SEM. Further experimental details are given in SI (Section S.3.B).

*b. SCF calculations.* SCF calculations help for a better understanding of the assembly of polyelectrolyte-surfactant complexes in the bulk, and of the adsorption of such complexes onto

Adsorption of PDADMAC+RL mixtures

solid surfaces [40, 41]. SCF are based on a minimization of a mean-field free-energy functional. For this purpose, we used the procedure introduced by Banerjee et al. [42, 43] (see Section S.4, SI).

# 3. Results and Discussion

## 3.1. Self-assembly of polyelectrolyte-glycolipids mixtures in solution

### 3.1.1. Experimental results

The apparent hydrodynamic diameter of the polyelectrolyte-surfactant complexes, and the monophasic or multiphasic character of the mixtures were evaluated for the characterization of the self-assembly of polyelectrolyte-surfactant mixtures in solution. It is worth mentioning that the poor solubility of mono-RL($C_{14}$) limits the study of their mixtures with PDADMAC to surfactant concentrations < 0.018 mM. Optical density dependences on the surfactant concentration, c, for the different studied polymer-surfactant mixtures are shown in Figure 1 (and Figure S.1, SI).

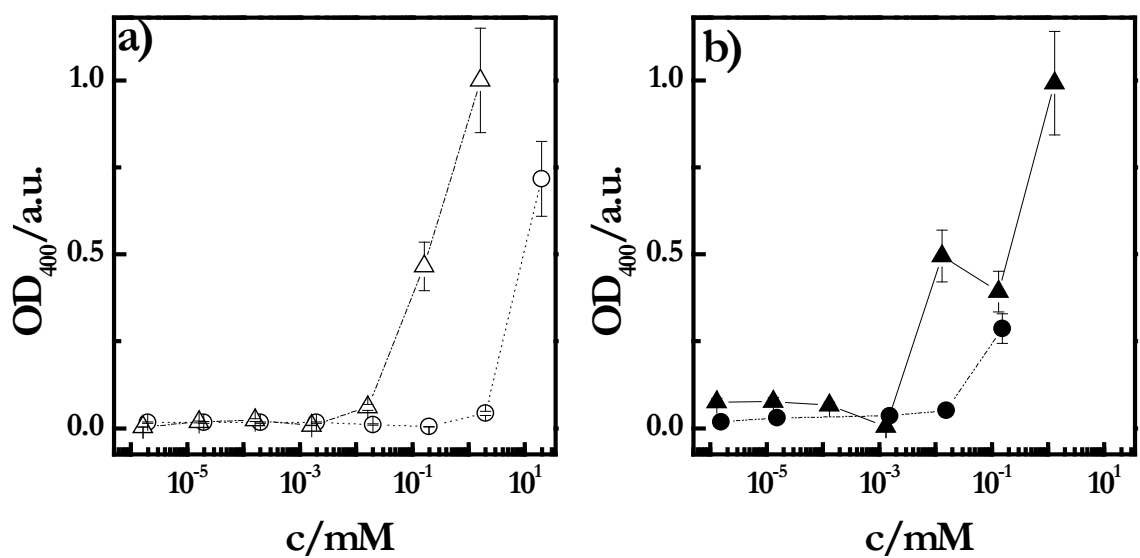

Adsorption of PDADMAC+RL mixtures

Figure 1. Dependences of the optical density measured at 400 nm on c for different PDADMAC – surfactant mixtures: (a) Effect of the length of the hydrophobic chain for mono-RL. (b) Effect of the length of the hydrophobic chain for di-RL. The symbols are referred to different samples: ○ mono-RL($C_{10}$), Δ mono-RL($C_{14}$), ● di-RL($C_{10}$), ▲ di-RL($C_{14}$) and ◊APG. The lines are guides for the eyes. The results correspond to PDADMAC-surfactant mixtures containing a fixed PDADMAC concentration of 0.5 wt% (pH = 5.6 and 40 mM of KCl concentration), and left to age for one week prior to measurement.

Transparent solutions (OD$_{400}$ ~ 0) were found for the lowest surfactant concentrations, independently of the considered surfactant. This is explained assuming that the ratio between the number of surfactant molecules (S) and monomers (P), S/P, is well below of 1, leading to the undercompensated complexes, where surfactant binding does not affect significantly neither the polymer charge nor the chain conformation. The increase of surfactant concentration leads to a sharp increase of the optical density, except for PDADMAC-APG mixtures (Figure S.1a, SI). This is a signature of the onset on the phase separation region, which should be considered as an unexpected result looking to the S/P ratio because even assuming the existence of a negligible amount of surfactant free in solution, it would be expected the formation of complexes with a significant excess of uncompensated monomer (S/P << 1), corresponding to an equilibrium one-phase region. However, two phase samples were found for surfactant concentrations more than one order of magnitude below the isoelectric point, i.e. the point in with a S/P ratio = 1 should be expected. Therefore, the onset in the phase separation may be only explained assuming the of formation kinetically trapped aggregates due to the Marangoni stresses created during the mixing process which are associated with a local excess of surfactant molecules. The kinetically trapped aggregates present a quasi-neutral inner core, even their external region remains charged. This later endows them of colloidal stability [44]. Adsorption of PDADMAC+RL mixtures

Such aggregates remain intact in the aqueous medium upon the final dilution of the mixtures, pushing the system to a two phase region for compositions that are far from the real equilibrium two-phase region. This scenario contrasts with the characteristic phase separation appearing in this systems associated with the lacked stability of the complexes formed in the isoelectric point [45-50]. The above scenario accounts for the behavior of the mixtures of PDADMAC with RL and SLES in which the electrostatic interactions are the main driving force for the assembly of the complexes. Figure 2 summarized the different steps of polyelectrolyte-surfactant complexation as the surfactant concentration increases

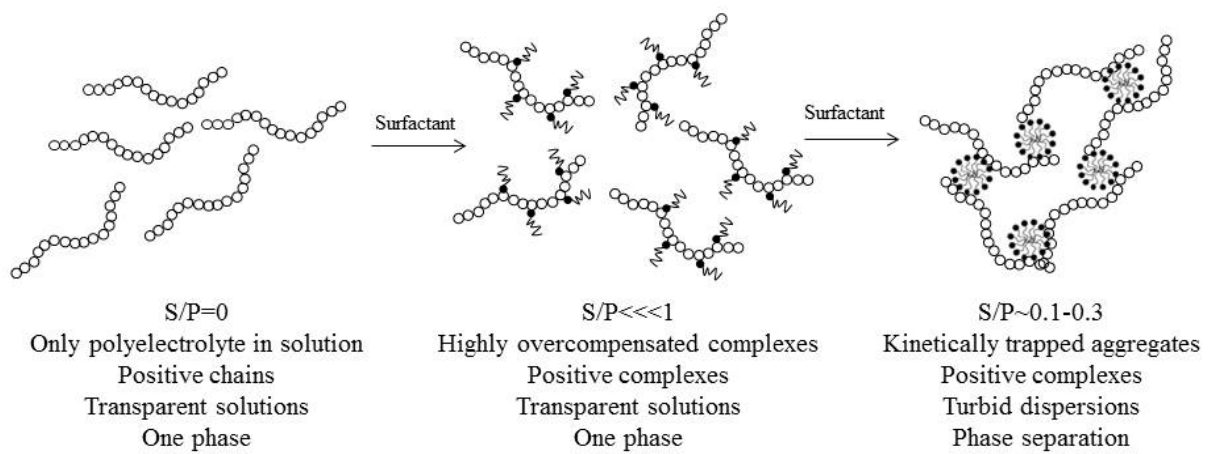

Figure 2. Schematic view the different steps involved in the polyelectrolyte-surfactant complexation with the increase of the surfactant concentration.

On the basis of the optical density is possible to understand the role of the surfactant hydrophobicity on the formation of macromolecular complexes. This is clear from mono-RL (Figure 1a) and di-RL (Figure 1b) results, where the sharp change of the turbidity appears for surfactant concentrations one order of magnitude lower when the length of the hydrocarbon chain increases from 10 to 14 carbon atoms. This is explained considering that the formation of micelles attached to the polymer chains appears from lower surfactant concentrations as the hydrophobicity of the surfactant increases, favoring the formation of bridge between polymer chains. The above discussion seems to be in apparent contradiction with the results obtained for surfactants with the same length of the hydrocarbon chain and different number of sugar rings in the hydrophilic head. In this case, di-RLs, which are expected to be more hydrophilic than mono-RLs, evidence higher increase of the turbidity with the surfactant concentration than mono-RLs (see SI, Figures S.1a and S.1b). This apparent discrepancy is explained considering the role of the head group size on the aggregation behavior of surfactant molecules.[51] Thus, the increase of the head group size decreases the aggregation number of the micelles or hemi-micelles bridging polymer chains, leading to the increase of the number of bridges between the chain and consequently to the rising of the turbidity for lower surfactant concentrations in di-RLs than in mono-RLs. For PDADMAC – APG mixtures, no increase of the turbidity was observed within the whole concentration range studied. This does not mean the absence of PDADMAC-APG complexes formation. In this case, the complexes are probably formed through hydrophobic interaction between the alkyl chain of the surfactant molecules, and the hydrophobic domains along the PDADMAC chains, without any role of the electrostatic interactions. The formation of PDADMAC-APG complexes is supported by the slight decrease of $D_{app}$ obtained for DLS for the highest surfactant concentrations (Figure S.2, SI).

DLS provides additional insights on the bulk aggregation of PDADMAC–surfactant mixtures. The apparent diffusion coefficient values (Figure S.2, SI), $D_{app}$, obtained from the analysis of the intensity auto-correlation functions obtained for the different polymer–surfactant mixtures give, assuming that the size is proportional to $D_{app}^{-1}$, qualitative information about the size of the aggregates. This allows explaining the decrease of $D_{app}$ as result of an increase of the average size of the complexes. Thus, DLS measurements shows that for low surfactant concentrations, the size of the aggregates remains rather constant, and almost independent on

Adsorption of PDADMAC+RL mixtures

the considered surfactant chosen and the surfactant concentration, appearing a sudden increase as the phase separation region is approached in agreement with the turbidity measurements (Figure 1 and Figure S.1 of SI).

## 3.1.2. SCF calculations

The binding isotherms obtained using SCF calculations provide additional insights in the self-assembly process of polyelectrolyte-surfactant in solution [42]. Figure 3 shows the binding isotherms obtained for the different PDADMAC-RL mixtures which provide information about the aggregation processes through the analysis of the aggregation number of the surfactant in the complexes, $g_S$, and the degree of polyelectrolyte-surfactant binding, $g_P$. The results show important differences in the complexes assembly when SLES is replaced by RL. For PDADMAC-SLES mixtures (Figure S.3a, SI), the aggregation number of surfactant in the complexes and the polyelectrolyte-surfactant binding increase with the chemical potential $\mu_P$, i.e. the polymer concentration, as result of the cooperative binding, i.e. binding of polymer to surfactant micelles increases progressively the surfactant aggregation number [43]. Close to the isoelectric point (S/P ~ 1), the spherical shape of the surfactant micelles start to be compromised and, the colloidal stability may be lost, i.e. when the binding overcomes the threshold value of 1. This is explained as result of the placing of the PDADMAC chains at the outer shell of the micelles as evidenced the radial volume fraction profiles, $\varphi$ (Figure S.4a, SI). For such systems, the loops and tails of PDADMAC allow the bridging between neighboring micelles, which prevents the colloidal stability.

When PDADMAC-RL mixtures is concerned, an anti-cooperative binding, i.e. the number of surfactant unit per micelle decreases with polymer binding, was found. Furthermore, it is worth mentioning that the differences in the interaction of RL (Figure 3) and APG (Figure S.3b, SI) with PDADMAC leads to the appearance of different scenario for the binding process in agreement with the experimental results. For PDADMAC-APG mixtures, the increase of Adsorption of PDADMAC+RL mixtures polymer concentration in the complexes does not lead to an isoelectric point, even the binding tend to increase as evidenced the increase of the composition ratio $f$ with $\mu_P$ (see inset in Figure S.3b, SI).

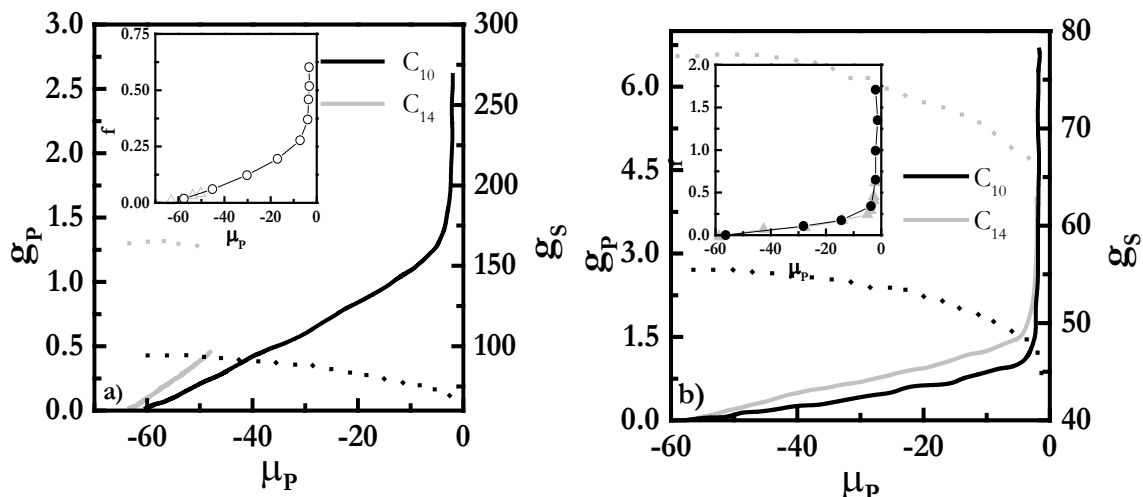

Figure 3. Binding isotherms for PDADMAC-Mono-RL (mono-RL($C_{10}$) and mono-RL($C_{14}$)) (a) and PDADMAC-Di-RL (di-RL($C_{10}$) and di-RL($C_{14}$)) (b) represented as the dependence on the PDADMAC chemical potential, $\mu_P$, of the aggregation number of the surfactant in the complexes, $g_S$, (dotted lines) and the degree of polyelectrolyte-surfactant binding, $g_P$, (solid lines). The inserted panels represent the dependence of the composition ratio $f$ in the complexes on $\mu_P$ for the same mixtures shown in the main panel. Note that $\mu_P$ is equivalent to the amount of polymer in the complexes and that the calculations were performed under spherical micelles configurations and fixed free energy $5k_BT$.

Additional insights on the binding process in PDADMAC-APG mixtures are obtained from the radial volume fraction profiles (Figure S.4b, SI) which shows that the polyelectrolyte always remains at the outer shell of the spherical micelles. However, the dimensionless charge density, $\xi(z) = q(z)/e$, profiles (inserted panel in Figure S.4b, SI) looks to be mostly positive, evidencing the absence of real charge compensation due to the neutral character of APG in agreement again with the experimental results. Therefore, it may be expected that Adsorption of PDADMAC+RL mixtures uncompensated charges remain available in PDADMAC-APG complexes, leading to complexes with are stable in a larger composition range than those involving charged surfactant. For high polymer concentrations, it would be expected a loss of stability of PDADMAC-APG complexes due to the depletion of surfactant molecules or the formation of bridges between polymer chains, with the hydrogen-bonding between head groups of APG being the driving force for the later. The calculated volume fraction profiles obtained for the particular case of $f = 0.44$ (inserted panel in Figure S.4d, SI) evidences that the PDADMAC-APG complexes are formed with the sugar rings remaining at their periphery, which leads to an association through hydrophobic interactions between the alkyl chains of the surfactant and the methyl groups surrounding the ammonium group along the polymer backbone, favoring the bridges through the hydrogen bond of the hydroxyl groups of different surfactant molecules.

The situation is slightly different for RL, both mono- and di-RL (Figure S.4b and S.4c, SI), even the binding is also non-cooperative. For RL, the existence of micelles which bind several PDADMAC chains was found. This results in the formation of multi-chain complexes. At low volume fraction of polymer, the dimensionless charge density of the complexes is similar to that found for PDADMAC-SLES (see insets in Figure S.4b and S.4c, SI), with an increasingly amount of surfactant molecules bound to the polymer chains. The increase of $g_P$ above 1 leads to the formation of complexes with positive charge. For the highest values of the volume fraction, it was found similar charge density profiles in PDADMAC-RL mixtures than when APG is the surfactant. This could be explained considering that the hydrogen bonds between the sugar rings may also participate in the stabilization of the complexes together with the hydrophobic and electrostatic interactions.

Paying attention to complexes containing mono-RL (Figure 3a), the SCF calculations confirm the poor solubility in water of the mono-RL with hydrophobic tail containing 14 carbon atoms, limiting the formation of spherical micelles, i.e. the sugar rings cannot compensate the increase
Adsorption of PDADMAC+RL mixtures

in hydrophobicity to form stable spherical micelles. For di-RL, the hydrophobicity does not affect significantly to the solubility of the complexes, and only a slight shift on the amount of surfactant needed for the complexation is found in agreement with the experimental results. The increase of the hydrophobicity of the di-RL leads to the formation of complexes with higher binding in agreement with the scenario expected from turbidity measurements. The comparison of the binding isotherms for mono- and di-RL (Figure 3a and 2b) for equivalent values of PDADMAC concentration evidences that the amount of surfactant needed for the formation of multichain complexes is lower in di-RL than in mono-RL, which agrees with the lower surfactant concentration needed for the onset in the phase separation region. This is associated with the decrease of the aggregation number of the micelles bound to the polymer chains.

## 3.2. Adsorption of polyelectrolyte-glycolipids mixtures onto model surfaces

### 3.2.1. Experimental results

The aggregation process of polymer–surfactant mixtures impacts significantly in the adsorption of the complexes onto solid surfaces. Figure 4 and Figure S.5 (see SI) show the acoustical thickness values, $h_{ac}$, obtained using QCM-D for the adsorption of PDADMAC–surfactant mixtures onto the solid surfaces.

$h_{ac}$ is substantially higher for mixtures containing RL than for PDADMAC–SLES mixtures, which can impact on the conditioning performance of RL in cosmetic formulations. In the low concentration range, i.e. when the solutions remain in the one-phase region, the adsorbed amount remains low and almost constant. In the onset of the phase separation region, the adsorbed amount starts to increase with surfactant concentration. This enhanced deposition in the onset of the phase separation is explained considering the combined action of the electrostatically driven deposition occurring as result of the opposite charge of complexes and surface, which is expected to play a role within the entire surfactant concentration range, and Adsorption of PDADMAC+RL mixtures the gravity driven sedimentation as result of the higher density of the kinetically trapped aggregated in relation to the solvent [52-54].

The analysis of the total adsorbed amount shows the following order: di-RL($C_{14}$) > mono-RL($C_{10}$) > di-RL($C_{10}$) > APG. No discussion is included for mixtures containing mono-RL($C_{14}$) due to its poor solubility. However, it is expected for these mixtures a maximal adsorption. The dependence of the adsorbed amount on the nature of the surfactant are associated with the hydrophobicity of the formed complexes, and consequently with the tendency to the phase separation and deposition onto the surface. It is worth mentioning that for PDADMAC–mono-RL($C_{10}$) mixtures a high total adsorbed amount was obtained from the lowest surfactant concentration. This may be due to the formation of heterogeneous layers trapping a high fraction of water (Note that QCM-D provides information of the total adsorbed amount deposited onto the quartz crystal, i.e. the solid materials and the solvent associated with the layer). On the basis of the results obtained using QCM-D it can be assumed that APG is not a good option for replacing SLES due to its limited adsorption. On the other side, mixtures containing PDADMAC and mono-RL($C_{14}$) are not recommended due to its limited solubility. Figure S.6 (see SI) shows the comparison of the $h_{ac}$ and the optical thickness, $h_{op}$, obtained by ellipsometry for layers of PDADMAC and mono-RL($C_{10}$), di-RL($C_{10}$) and di-RL($C_{14}$) (Note that $h_{ac}$ is higher than $h_{op}$ because whereas QCM-D does not distinguish between complexes adsorption and water associated with the layers ellipsometry gives information only about the adsorbed complexes [55-57]). Ellipsometry confirms the above described scenario for the adsorption of polyelectrolyte – surfactant complexes (Figure S.6 in SI. Note: the study of the adsorption onto the solid surface using ellipsometry was not possible for turbid samples).

Adsorption of PDADMAC+RL mixtures

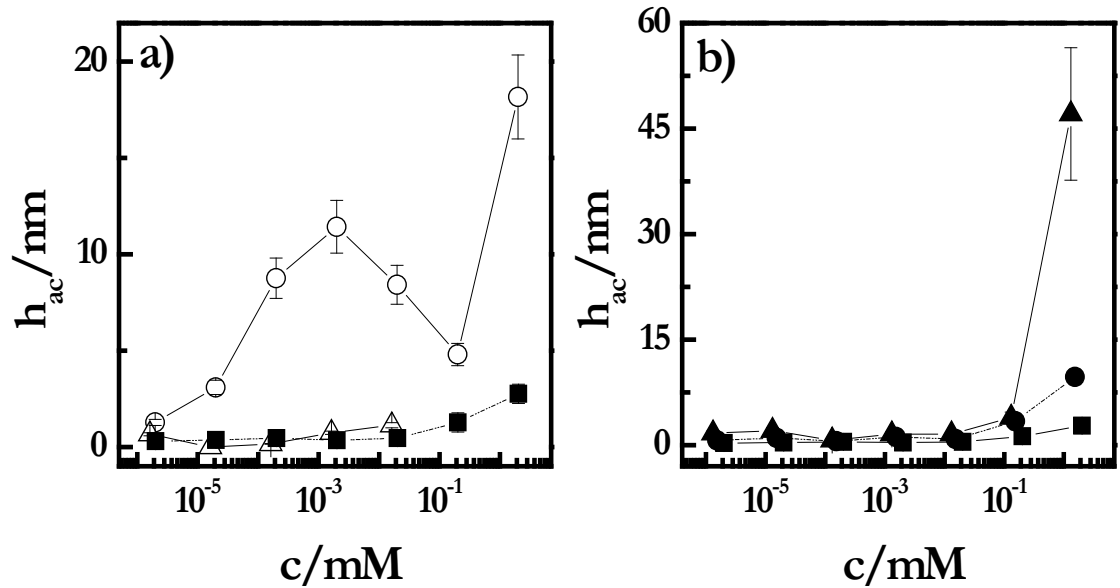

Figure 4. $h_{ac}$ dependences on c for the different PDADMAC– surfactant mixtures studied: (a) Effect of the length of the hydrophobic chain for mono-RL and (b) Effect of the length of the hydrophobic chain for di-RL. The symbols are referred to different samples: ○ mono-RL($C_{10}$), △ mono-RL($C_{14}$), ● di-RL($C_{10}$), ▲ di-RL($C_{14}$) and ◇APG. In all the plots, the data corresponding to samples PDADMAC – SLES are included (■).The lines are guides for the eyes. The results correspond to PDADMAC-surfactant mixtures containing a fixed PDADMAC concentration of 0.5 wt% (pH = 5.6 and 40 mM of KCl concentration), and left to age for one week prior to measurement.

Ellipsometry shows that for most of the studied polyelectrolyte–surfactant mixtures, the effective surface concentration of solid material is rather limited, and only for di-RL($C_{14}$), the adsorption obtained using ellipsometry is comparable to that found by QCM-D. Therefore, it is possible to assume that most of the layers are highly hydrated, whereas the complexes formed by di-RL($C_{14}$) and PDADMAC are more shrank presenting less amount of associated water, leading to the formation of more dense layers. This is confirmed comparing ellipsometry and QCM-D results to calculate the water content, $X_w$, of the layers, whereas PDADMAC-di-Adsorption of PDADMAC+RL mixtures

RL(C$_{14}$) evidences values between 10-30 % in the whole concentration range, the water included within the layer remains rather constant around 80% for mixtures containing the other RL, and only close to the phase separation a sharp decrease of the water content of the layers was found. The lower solubility of the complexes of PDADMAC – di-RL(C$_{14}$) can explain the lower average value of the water content found within the entire surfactant concentration range in relation to the values obtained for the other mixtures. At this point, it is necessary to mention that the conditioning performance of formulation is based on an intricate balance between the maximization of deposition occurring due to the precipitation of the complexes onto the hair fiber and the friction properties of the coated fibers associated with the water content of the layers.[3, 58, 59] The results of ellipsometry (see Figure S.6, SI) related to the solid adsorbed amount onto solid surfaces, show that all the mixtures present an almost negligible adsorption for the lowest surfactant concentrations, increasing as the onset of the phase separation region is approached. The above discussion allows one to assume that on the basis of the adsorbed amount, PDADMAC - di-RL(C$_{14}$) mixtures increases in one order of magnitude the deposited amount in relation to mixtures with SLES for composition in the onset of the phase separation region. However, the water content (see Figure S.4e in SI for a comparison of the mixtures) of the layers is rather similar to that found for the PDADMAC–SLES layers, affecting to its lubricating properties. The case of mixtures containing di-RL(C$_{10}$) is just the opposite, the water content is high whereas the adsorbed amount is even lower than that found for PDADMAC–SLES mixtures. On the basis of the above results, the substitution of SLES by di-RL does not involve a significant improvement of the deposit properties. The mixtures of PDADMAC and APG do not introduce any significant difference in relation to the characteristic of layer of PDADMAC – SLES mixtures. The situation is different when mixtures of PDADMAC and mono-RL(C$_{10}$) are analyzed. In this latter case, the adsorption is only twice higher than for mixtures of PDADMAC and SLES, and the water content also

Adsorption of PDADMAC+RL mixtures

increases in relation to that found for the mixtures containing SLES, making of mono-RL($C_{10}$) a good alternative for replacing SLES.

To deepen on the potential application of mixtures of PDADMAC and mono-RL($C_{10}$) in conditioning formulations it is needed to analyze the topography of the layers adsorbed onto the solid surfaces. Taking into consideration the limited adsorption of the mixtures containing surfactant concentrations below $10^{-1}$ mM, a focus was made on the highest concentrations. Figure 5 shows AFM images obtained in Tapping mode and the height profiles for samples of PDADMAC - mono-RL($C_{10}$) mixtures with concentrations of $10^{-1}$ and 1 mM.

The AFM images show a high coverage of the surface by the mixtures that may be considered an additional advantage in the conditioning process. The AFM images show the densification of the layers as the surfactant concentration increases, appearing lower density of holes in the layer adsorbed from solution with surfactant concentration around 1 mM than in that with concentration $10^{-1}$ mM. This agrees well with the fact that the precipitation and collapse of the aggregates onto the solid surfaces increases as the phase separation is approached. Thus, the adsorption is favored and consequently, should bring an improved conditioning effect. It is worth mentioning that AFM images shows the existence of layers containing two types of structural patterns, (i) large domains with lower contrasts, and (ii) brighter dots. This is explained considering the formation of layers with presence of compact complexes deposited by electrostatically-driven adsorption (lower contrast regions) and kinetically-trapped deposited by gravitational transport [53, 54].

Adsorption of PDADMAC+RL mixtures

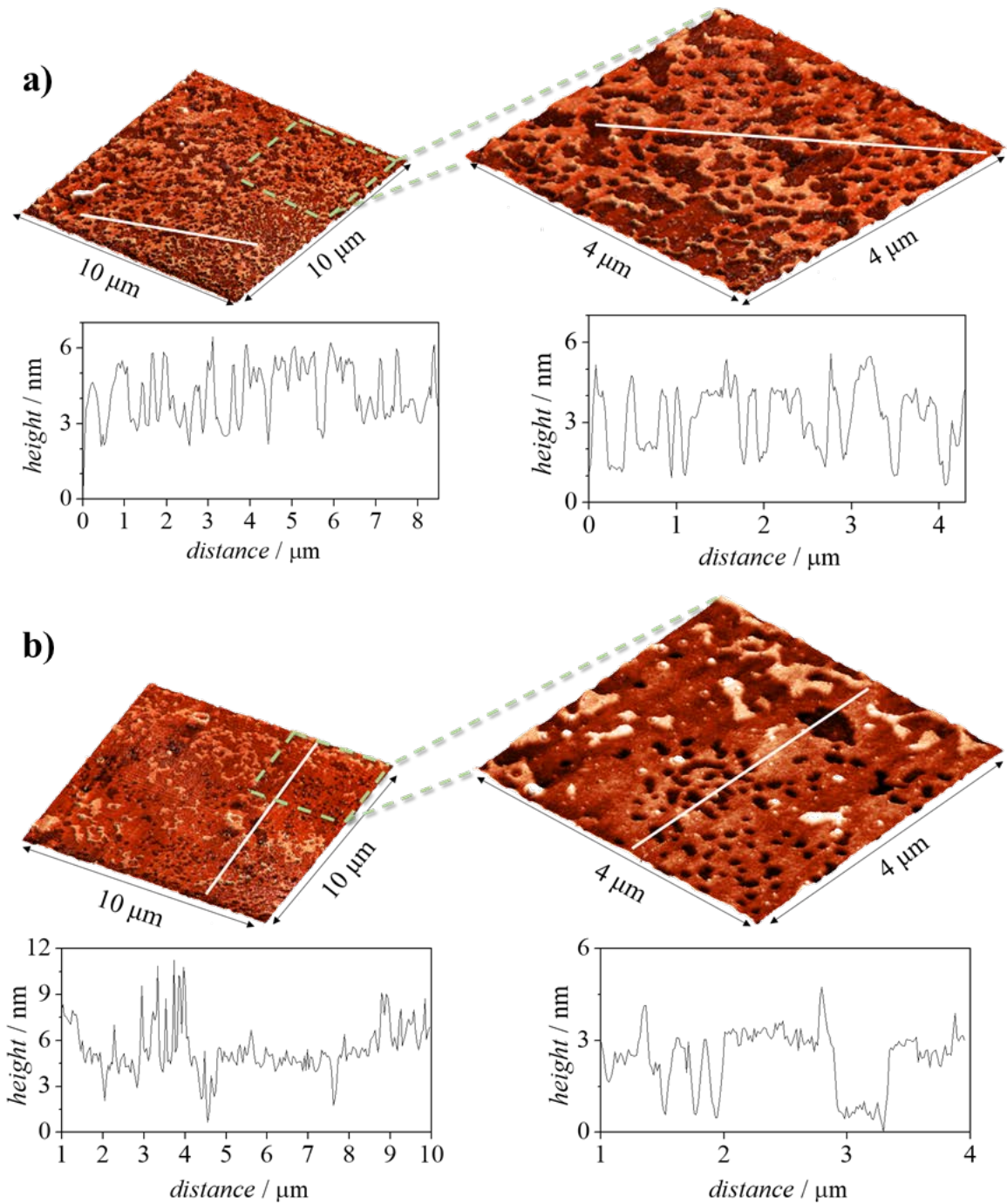

Figure 5. AFM images and their corresponding height profiles for PDADMAC - mono-RL(C$_{10}$) mixtures with surfactant concentration $10^{-1}$ mM (a) and 1 mM (b) adsorbed onto the solid surfaces. Two different AFM images with different level of magnification (10x10 and 4x4 $\mu$m$^2$) are shown for each sample. The results correspond to PDADMAC-surfactant mixtures

Adsorption of PDADMAC+RL mixtures

containing a fixed PDADMAC concentration of 0.5 wt% (pH = 5.6 and 40 mM of KCl concentration), and left to age for one week prior to measurement.

### 3.2.2. SCF calculations

SCF calculations also give information related to the adsorption of polyelectrolyte-surfactant onto charged surfaces at the same fixed chemical potential used for bulk calculations. This is possible through the ratio between the adsorbed amount, $\theta_{ads}$, of glycolipids and that of SLES in mixtures with low values of $f$, and the same ratio for the adsorbed amount of PDADMAC in the same mixtures (Figure S.7, SI). These calculations were performed for surface with different hydrophobicity, $\sigma$. Note that hydrophilic surfaces ($\sigma$=0) corresponds to bleached or damaged hair surfaces with very low density of hydrophobic alkyl groups, whereas the increase of the amount of hydrophobic groups, i.e. increasing $\sigma$, should lead to a behavior closer to the surface of a less damaged hair fibers.

The behavior of PDADMAC-APG and PDADMAS-SLES mixtures (Figure S.7, SI) is similar, with the adsorbed amount of surfactant being equivalent over the whole range of hydrophobicities. The increase of $f$ leads to a decrease on the adsorbed density of APG on the bare surface in relation to the adsorption of SLES. The similarity in the behavior of the mixtures containing SLES and APG agrees with the experimental results. The amount of adsorbed PDADMAC is always smaller for PDADMAC-APG mixtures than for PDADMAC-SLES one, dropping to zero with the increase of the surface hydrophobicity. This is due to the absence of charge in APG which leads to an enhanced APG adsorption through the alkyl tails with the increase of the surface hydrophobicity. However, this does not provide a favorable environment for the electrostatically driven adsorption of PDADMAC. On the contrary, the adsorption of SLES through its hydrophobic tail place the charged groups protruding to the Adsorption of PDADMAC+RL mixtures aqueous phase, and this negatively charged environment provides the bases for PDADMAC adsorption. The analysis of the volume fractions perpendicular to the surface for the different species helps to understand the deposition mechanism onto hydrophilic surfaces with negative charge (Figure 6a). The results show that PDADMAC adsorbs through electrostatics interaction in mixtures with both APG and SLES. However, the absence of charges in APG hinders the co-adsorption of PDADMAC through surfactant-mediated adsorption.

The peak found in Figure S.7 (see SI) in the $\sigma$ range 0.08-0.15 indicates the cohabitation of two different adsorption regime: (i) electrostatically driven deposition of PDADMAC onto the surfaces through its charged monomers, which leads to the formation of the polymer brush allowing a co-adsorption of surfactant when they are charged in agreement with the results in Figure 6a and 5b, and the experimental results, and (ii) co-adsorption of the surfactant and the polymer through weak hydrophobic interactions between the alkyl chain of the surfactant and the hydrophobic domain in PDADMAC backbone. Furthermore, the appearance of an additional adsorption regime is possible when the alkyl tails of surfactants first adsorb onto the hydrophobic regions of the surface, providing a negatively charged layer (for RL) that promotes the adsorption of PDADMAC. Further increases of the hydrophobicity of the surface hinder the electrostatic-driven interaction of PDADMAC with the surface, leading to the enhancement of the second regime.

Adsorption of PDADMAC+RL mixtures

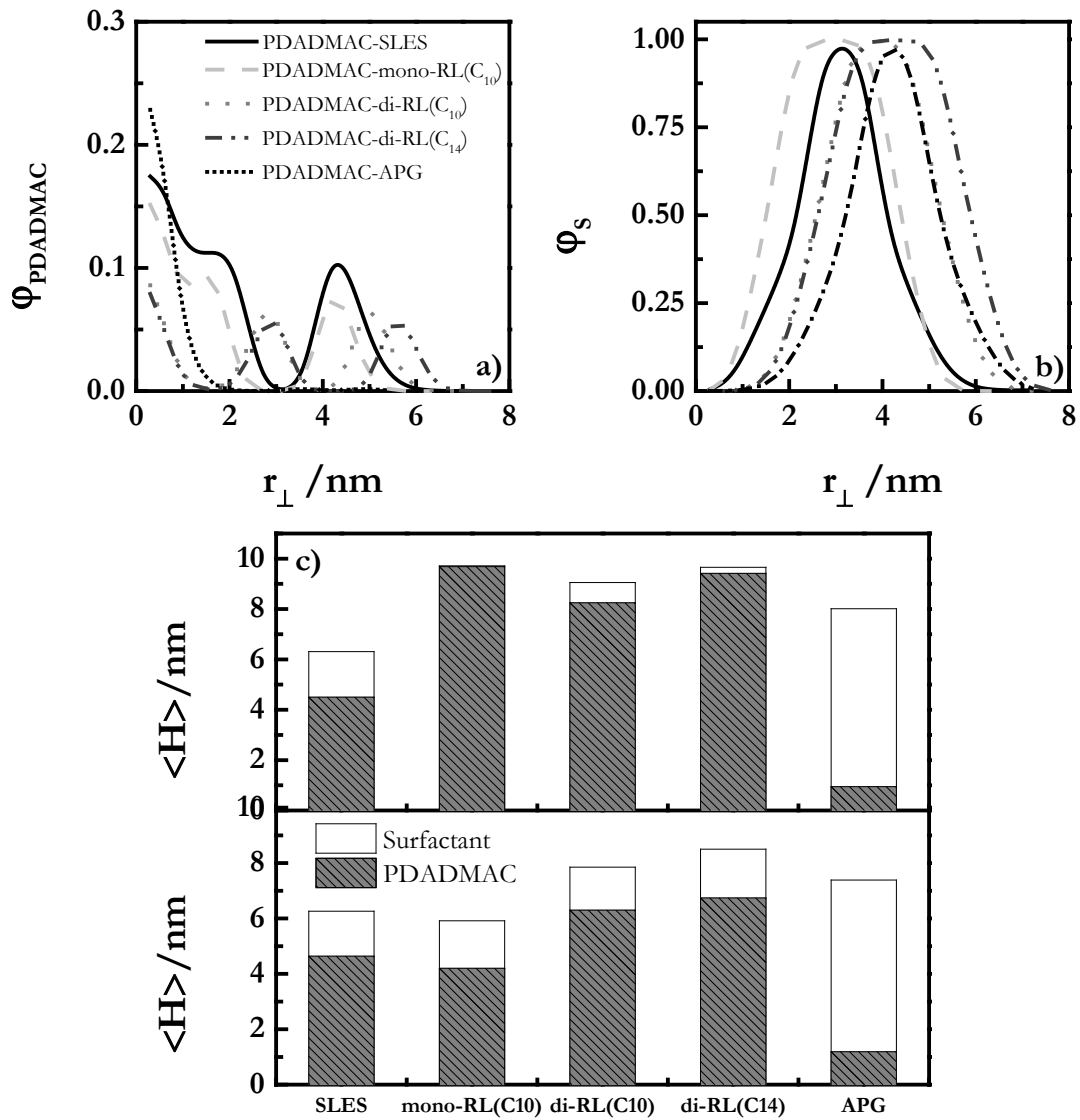

Figure 6. (a) Volume fraction profiles of PDADMAC, $\varphi_{PDADMAC}$, in the direction perpendicular to the surface for the adsorbed layers onto bare negatively charged surfaces obtained for the different PDADMAC-surfactant mixtures. (b) Volume fraction profiles of surfactant, $\varphi_s$, in the direction perpendicular to the surface for the adsorbed layers onto bare negatively charged surfaces obtained for the different PDADMAC-surfactant mixtures. Notice that the legend is the same that for panel a. (c) Average height $\langle H \rangle$ of each species in the deposits for the different binary systems onto purely hydrophilic negatively charged bare surface obtained at two different ratio $f = 5 \cdot 10^{-4}$ (top) and $f = 3 \cdot 10^{-2}$ (bottom).

Adsorption of PDADMAC+RL mixtures

The results show an enhancement of the adsorption of PDADMAC–mono-RL($C_{10}$) in relation to PDADMAC-APG mixtures (Figure S.7, SI) over the whole range of hydrophobicity degree of the surface. It is worth mentioning that the amount of PDADMAC is comparable to that found for PDADMAC-SLES mixtures, which is mainly the result of the size of the groups of the glycolipids that disfavors PDADMAC adsorption. This is confirmed for the volume fraction profiles shown in Figure 6a and 5b, with the brush formed by PDADMAC presenting less importance in the PDADMAC–mono-RL($C_{10}$) mixture than in PDADMAC–SLES one. The comparison of the volume fraction profiles obtained for mixtures containing mono-RL($C_{10}$) and those containing di-RL evidences some characteristic features depending on the surfactant. For di-RL, the adsorption of PDADMAC occurs less directly onto the surface. However, the brush for surfactant co-adsorption is longer which results from the larger size of the head groups. This forces the development of a thicker polymer brush as result of the binding of the carboxylic groups of the surfactants to the polymer. At equivalent composition fraction, $f$, and hydrophobic group $C_{10}$, the difference in head group size between mono- and di-RL may be considered the only reason that changes the adsorption profiles and the thickness of the adsorbed layer. As a counterpart, the thickness of the adsorbed layer of surfactant is larger for RL than for SLES as shown the volume fraction profiles of surfactant in Figure 6b, and the more hydrophobic the RL, the thicker it is. This could have a positively impact on the cosmetic application of the mixtures.

Deepening on the behavior of di-RL presenting different hydrophobicity, it is found that the increase of the length of the alkyl chain of the di-RL enhances the deposition of surfactant over the whole range of hydrophobicity of the surfaces (Figure S.7c and S.7d, SI). However, the deposited PDADMAC amount in the layers is worsened with the increase of the surfactant hydrophobicity.

Adsorption of PDADMAC+RL mixtures

The SCF method allows also calculating the average height of each adsorbed species (Figure 6c), $\langle H \rangle = 2 \int dr\, r\, \varphi(r) / \int dr\, \varphi(r)$ with φ and r being the volume fraction profile and r the direction perpendicular to the surface (The factor 2 ensures that $\langle H \rangle$ reduces to the layer thickness for a step density profile). The average heights of deposits for PDADMAC-SLES and PDADMAC-APG systems is roughly independent of the composition fraction, *f*. Noticeable differences occur for RL, more pronounced for mono-RL($C_{10}$). A decrease of roughly 2-3 nm for di-RL($C_{10}$ and $C_{14}$) thickness layers and around 5 nm for PDADMAC, whereas the decrease for PDADMAC-mono-RL($C_{10}$) of around 5 nm was found for both PDADMAC and RL between the two different composition ratios. Apparently, there is interplay between RL and polymers as composition ratio, *f*, increases that shrinks the adsorbed layers. In any case, di-RL provides a thicker layer than SLES, APG and mono-RL($C_{10}$), as was observed experimentally, whereas mono-RLs could also be a good candidate to replace SLES in formulations, as it offers a similar deposition.

### 3.3. Adsorption of polyelectrolyte-glycolipid mixtures onto hair fibers

The adsorption of polyelectrolyte – surfactant mixtures onto bleached hair fibers has been studied by SEM. For the sake of example, Figure S.8 (see SI) shows a set of images of a hair fiber before and after a treatment with a PDADMAC-RL mixture, evidencing that hair fibers are significantly coated after immersion in a polymer-surfactant solutions, with time being an important parameter to enhance the deposition (Further details are given in Section S.4.E in SI). Deepening on the adsorption, studies on the evolution of the adsorption to the inner region of the hair were performed. For this purpose, the analysis of sections of fibers before and after the treatment was performed using SEM. Figure 7 shows the image of a transversal cut of an untreated hair fiber.

Adsorption of PDADMAC+RL mixtures

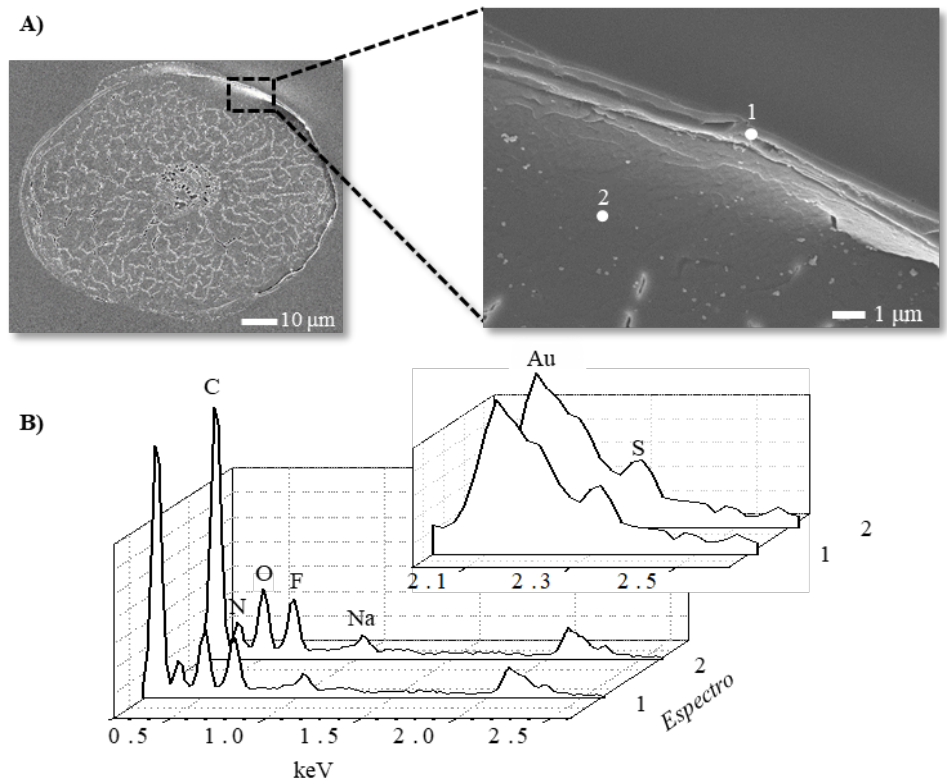

Figure 7. (a) Image of a transversal cut of an untreated hair fiber and the enlargement of the most external region. The number indicates the region in which EDS analysis were performed. (b) EDS spectra for the two regions evidenced in panel a. The results correspond to PDADMAC-surfactant mixtures containing a fixed PDADMAC concentration of 0.5 wt% (pH = 5.6 and 40 mM of KCl concentration), and left to age for one week prior to measurement.

The SEM image shows the layered structure of hair, with the cuticle being the most external layer. Continuing the analysis from the external region to the inner one, the appearance of the cortex and medulla is observed [26]. The EDS analysis (Figure 7b) is in good agreement with the above discussed for the analysis of the untreated cuticle (region 1). Notice the presence of Na and F in the inner region of the hair (region 2). The origin of these components is not clear. However, it can be associated with a bioaccumulation phenomenon as was previously

Adsorption of PDADMAC+RL mixtures

described in the literature [60, 61]. Figure 8 shows a similar set of images to those in Figure 7 for a fiber treated with a PDADMAC-RL mixture.

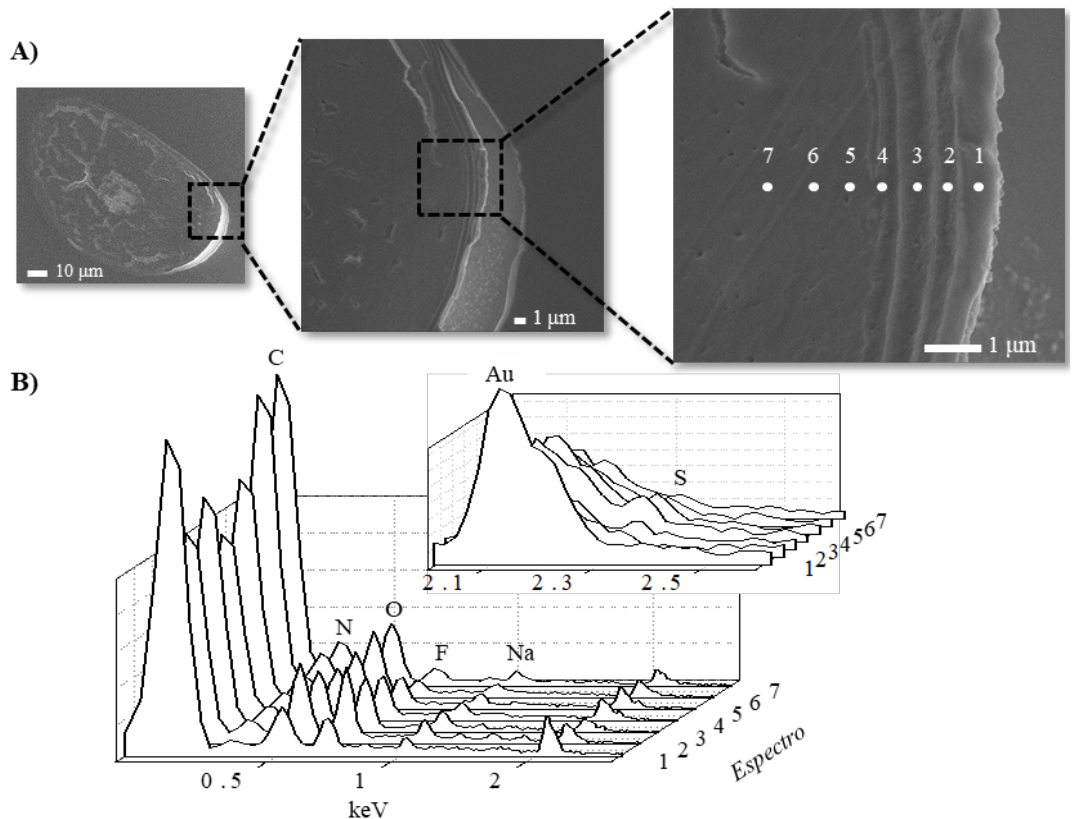

Figure 8. (a) Image of a transversal cut of a hair fiber treated with a PDADMAC - PDADMAC - mono-RL($C_{10}$) solution with concentration of concentration 0.1 mM and two enlargement of the most external. The numbers indicate the region in which EDS analysis was performed. (b) EDS spectra for the different regions evidences in panel a. The results correspond to PDADMAC-surfactant mixtures containing a fixed PDADMAC concentration of 0.5 wt% (pH = 5.6 and 40 mM of KCl concentration), and left to age for one week prior to measurement.

The SEM images show the efficiency of the deposition in the most external region of the hair fiber. However, SEM images do not give a confirmation of the penetration of the mixture to the inner region of the fiber. EDS spectra for treated hair are in good agreement with those shown in Figure 7b. However, the detailed analysis of the whole set of EDS spectra evidences Adsorption of PDADMAC+RL mixtures the presence of S in the inner regions of the hair, giving indication of a limited penetration of polymer-surfactant mixture beyond the most external region. This suggest that conditioning formulations works mainly following a surface mediated processm with the adsorption onto the cuticle playing a higher role than any possible absorption to the inner region of the fibers. This also validates, at least qualitatively, the use of solid model surfaces for obtaining insights on the physico-chemical bases of the conditioning properties of polyelectrolyte - surfactant mixtures.

## 4. Conclusions

The conditioning process is mainly governed by the deposition of polymer and surfactants onto damaged hair fiber with negative surface charge. Hence, the study of the adsorption of polymer–surfactant mixtures onto solid surfaces with a negative charge similar to that of damaged hair fibers should be a good model for understanding the physico-chemical bases underlying the conditioning process. This work studied the effect of the replacement of SLES for glycolipids in the performance of their mixtures with PDADMAC in the adsorption onto solid surfaces, seeking for alternatives to SLES in polyelectrolyte-surfactant mixtures of cosmetic interest. A good understanding of the deposition process of polyelectrolyte-surfactant mixtures requires information related to the aggregation behavior of polymers and surfactant in bulk, which has been found to be strongly correlated to the hydrophilic-lipophilic balance of the surfactants. Such balance is governed by an intricate balance between the effect of the number of rhamnose rings in the hydrophilic head and the length of hydrocarbon chains. The results have evidenced that the increase of the surfactant concentration drives the system to the phase separation due to the formation of kinetically-trapped aggregates far of the isoelectric point, with this playing a key role in the conditioning process. The different mixtures showed a qualitative similar behavior in relation to their bulk aggregation and adsorption onto negatively charged surfaces. However, for most of the mixtures containing glycolipids, except those with APG, a significant increase on the adsorption was found in relation to the PDADMAC–SLES mixture. Therefore, RL increase the efficiency of the formulations in relation to those including SLES. However, the analysis of the hydration degree of the layers reduced the number of possibilities to replace PDADMAC – SLES mixtures to PDADMAC – mono-RL(C$_{10}$) mixtures, which lead to a high degree of surface coverage, enhancing the deposition and hydration degree of the layers. On the basis of the here obtained results may be proposed the use of RL structures with short alkyl chains length for replacement SLES from their mixtures with PDADMAC.

**Acknowledgements**

This work was funded in part by L'Oréal S.A and by MINECO under grant CTQ2016-78895-R. We are grateful to C.A.I. Espectroscopía y Correlación from the UCM for the use of their facilities. We thank Céline Farcet (L'Oréal) for supporting this study.

**Competing financial interests**

G.S.L, L.T.C., K.R. and F.L. are employees of L'Oréal. The rest of authors are from Complutense University, having received financial contributions from L'Oréal for this study.

**References**

[1] M.J. Rosen, Surfactants in Emerging Technologies, CRC Press, New York, United States of America, 1987.
[2] P. Fisk, Chemical Risk Assessment: A Manual for REACH, Wiley-Blackwell, Hoboken, United States of America, 2013.

Adsorption of PDADMAC+RL mixtures

[3] G.S. Luengo, A. Galliano, C. Dubief, Aqueous Lubrication in Cosmetic, in: N.D. Spencer (Ed.) Aqueous Lubrication. Natural and Biomimetic Approaches, World Scientific Publishing Co. Pte. Ltd., Singapore, Singapore, 2014, pp. pp 103-144. doi: 110.1142/9789814313773_9789814310004.

[4] E.D. Goddard, J.V. Gruber, Principles of Polymer Science and Technology in Cosmetics and Personal Care, Marcel Dekker, Inc., Basel, Switzerland, 1999.

[5] C.G. Gebelein, T.C. Cheng, V.C. Yang, Cosmetic and Pharmaceutical Applications of Polymers, Plenum Press, New York, United States of America, 1991.

[6] S. Llamas, E. Guzmán, F. Ortega, N. Baghdadli, C. Cazeneuve, R.G. Rubio, G.S. Luengo, Adsorption of polyelectrolytes and polyelectrolytes-surfactant mixtures at surfaces: a physico-chemical approach to a cosmetic challenge, Adv. Colloid Interface Sci., 222 (2015) 461-487. doi: 410.1016/j.cis.2014.1005.1007.

[7] E. Guzmán, V.S. Miguel, C. Perinado, F. Ortega, R.G. Rubio, Polyelectrolyte Multilayers Containing Triblock Copolymers of Different Charge Ratio, Langmuir, 26 (2010) 11494-11502. doi: 11410.11021/la101043z.

[8] K.K. Sekhon Randhawa, P.K.S.M. Rahman, Rhamnolipid biosurfactants—past, present, and future scenario of global market, Front. Microbiol., 5 (2014) 454. doi: 410.3389/fmicb.2014.00454.

[9] P.K.S.M. Rahman, E. Gakpe, Production, characterization and applications of biosurfactants-review, Biotechnology 7(2008) 360-370. doi: 310.3923/biotech.2008.3360.3370.

[10] M. Pacwa-Płociniczak, G.A. Płaza, Z. Piotrowska-Seget, S.S. Cameotra, Environmental applications of biosurfactants: recent advances, Int. J. Mol. Sci., 12 (2011) 633-654. doi: 610.3390/ijms12010633.

[11] M.J. Brown, Biosurfactants for cosmetic applications, Int. J. Cosmet. Sci., 13 (1991) 61-64.

[12] J.D. Desai, I.M. Banat, Microbial production of surfactants and their commercial potential, Microbiol. Mol. Biol. Rev., 61 (1997) 47-64.

[13] A. Varvaresou, K. Iakovou, Biosurfactants in cosmetics and biopharmaceuticals, Lett. Appl. Microbiol., 61 (2015) 214-223. doi: 210.1111/lam.12440.

[14] Y. Zhou, S. Harne, S. Amin, Optimization of the Surface Activity of Biosurfactant–Surfactant Mixtures, J. Cosmet. Sci., 70 (2019) 127-136.

[15] L. Xu, S. Amin, Microrheological study of ternary surfactant-biosurfactant mixtures, Int. J. Cosmetic Sci., 41 (2019) 364-370. doi: 310.1111/ics.12541.

[16] J. Penfold, R.K. Thomas, H.-H. Shen, Adsorption and self-assembly of biosurfactants studied by neutron reflectivity and small angle neutron scattering: glycolipids, lipopeptides and proteins, Soft Matter, 8 (2012) 578-591. doi: 510.1039/c1031sm06304a.

[17] N. Lourith, M. Kanlayavattanakul, Natural surfactants used in cosmetics: glycolipids, Int. J. Cosmetic Sci., 31 (2009) 255-261.

[18] K. Muthusamy, S. Gopalakrishnan, T.K. Ravi, P. Sivachidambaram, Biosurfactants: Properties, commercial production and application Curr. Sci., 94 (2008) 736-747.

[19] M.M. Rieger, L.D. Rhein, Surfactants in Cosmetics, Marcel Dekker, New York, United States of America, 1997.

[20] E.W. Flick, Cosmetic and Toiletry Formulations, Noyer Publication-William Andrew Publishing, LLC, Norwich-New York, USA, 1999.

[21] A.M. Abdel-Mawgoud, F. Lépine, E. Déziel, Rhamnolipids: diversity of structures, microbial origins and roles, Appl. Microbiol. Biotechnol., 86 (2010) 1323-1336. doi: 1310.1007/s00253-00010-02498-00252.

Adsorption of PDADMAC+RL mixtures

[22] E. Guzmán, F. Ortega, N. Baghdadli, C. Cazeneuve, G.S. Luengo, R.G. Rubio, Adsorption of Conditioning Polymers on Solid Substrates with Different Charge Density, ACS Appl. Mat. Interfaces, 3 (2011) 3181-3188. doi: 3110.1021/am200671m.

[23] E. Guzmán, F. Ortega, M.G. Prolongo, V.M. Starov, R.G. Rubio, Influence of the molecular architecture on the adsorption onto solid surfaces: comb-like polymers, Phys. Chem. Chem. Phys., 13 (2011) 16416-16423. doi: 16410.11039/c16411cp21967g

[24] S. Llamas, E. Guzmán, N. Baghdadli, F. Ortega, C. Cazeneuve, R.G. Rubio, G.S. Luengo, Adsorption of poly(diallyldimethylammonium chloride)—sodium methyl-cocoyl-taurate complexes onto solid surfaces, Colloids Surf. A, 505 (2016) 150-157. doi: 110.1016/j.colsurfa.2016.1003.1003.

[25] M. Korte, S. Akari, H. Kühn, N. Baghdadli, H. Möhwald, G.S. Luengo, Distribution and Localization of Hydrophobic and Ionic Chemical Groups at the Surface of Bleached Human Hair Fibers, Langmuir, 30 (2014) 12124-12129. doi: 12110.11021/la500461y.

[26] C.R. Robbins, Chemical and Physical Behavior of Human Hair, Springer Berlin, Germany, 2012.

[27] S. Breakspear, J.R. Smith, G. Luengo, Effect of the covalently linked fatty acid 18-MEA on the nanotribology of hair's outermost surface, J. Struct. Biol., 149 (2005) 235–242. doi:210.1016/j.jsb.2004.1010.1003.

[28] N. Baghdadli, G.S. Luengo, A closer look at the complex hydrophilic/hydrophobic interactions forces at the human hair surface, J. Phys. Chem. B, 100 (2008) 052034. doi:052010.051088/051742-056596/052100/052035/052034.

[29] E. Guzmán, F. Ortega, N. Baghdadli, G.S. Luengo, R.G. Rubio, Effect of the molecular structure on the adsorption of conditioning polyelectrolytes on solid substrates, Colloids Surf. A, 375 (2011) 209–218. doi:210.1016/j.colsurfa.2010.1012.1012.

[30] W. Xiong, Z. Zeng, X. Li, G. Zeng, R. Xiao, Z. Yang, H. Xu, H. Chen, J. Cao, C. Zhou, L. Qin, Ni-doped MIL-53(Fe) nanoparticles for optimized doxycycline removal by using response surface methodology from aqueous solution, Chemosphere, 232 (2019) 186-194. doi: 110.1016/j.chemosphere.2019.1005.1184.

[31] W. Xiong, G. Zeng, Z. Yang, Y. Zhou, C. Zhang, M. Cheng, Y. Liu, L. Hu, J. Wan, C. Zhou, R. Xu, X. Li, Adsorption of tetracycline antibiotics from aqueous solutions on nanocomposite multi-walled carbon nanotube functionalized MIL-53(Fe) as new adsorbent, Sci. Total Environ., 627 (2018) 235-244. doi: 210.1016/j.scitotenv.2018.1001.1249.

[32] J. Cao, B. Xu, H. Lin, B. Luo, S. Chen, Chemical etching preparation of BiOI/BiOBr heterostructures with enhanced photocatalytic properties for organic dye removal, Chem. Eng. J., 185-186 (2012) 91-99. doi: 10.1016/j.cej.2012.1001.1035.

[33] R.S. Reis, A.G. Pereira, B.C. Neves, D.M.G. Freire, Gene regulation of rhamnolipid production in Pseudomonas aeruginosa – A review, Bioresour. Technol. , 102 (2011) 6377-6384. doi:6310.1016/j.biortech.2011.6303.6074.

[34] T. Sarachat, O. Pornsunthorntawee, S. Chavadej, R. Rujiravanit, Purification and concentration of a rhamnolipid biosurfactant produced by Pseudomonas aeruginosa SP4 using foam fractionation, Bioresour. Technol., 101 (2010) 324-330. doi: 310.1016/j.biortech.2009.1008.1012.

[35] S. George, K. Jayachandran, Production and characterization of rhamnolipid biosurfactant from waste frying coconut oil using a novel Pseudomonas aeruginosa D., J. Appl. Microbiol., 114 (2013) 373-383. doi: 310.1111/jam.12069.

[36] M.V. Voinova, M. Rodahl, M. Jonson, B. Kasemo, Viscoelastic acoustic responde of layered polymer films at fluid-solid interfaces: continuum mechanics approach, Phys. Scr., 59 (1999) 391-396. 310.1238/Physica.Regular.1059a00391.

Adsorption of PDADMAC+RL mixtures

[37] R.M.A. Azzam, N.M. Bashara, Ellipsometry and Polarized Light, Elsevier, North-Holland, The Netherlands, 1987.

[38] E. Guzmán, H. Ritacco, J.E.F. Rubio, R.G. Rubio, F. Ortega, Salt-induced changes in the growth of polyelectrolyte layers of poly(diallyldimethylammoniumchloride) and poly(4-styrene sulfonate of sodium), Soft Matter, 5 (2009) 2130-2142. doi: 2110.1039/b901193e.

[39] P. Nestler, C. Helm, Determination of refractive index and layer thickness of nm-thin films via ellipsometry, Opt. Express, 25 (2017) 301052. doi: 301010.301364/OE.301025.027077.

[40] J.M.H.M. Scheujtens, G.J. Fleer, Statistical theory of the adsorption of interacting chain molecules. 1. Partition function, segment density distribution, and adsorption isotherms, J. Phys. Chem., 83 (1979) 1619-1635. doi: 1610.1021/j100475a100012

[41] O.A. Evers, J.M.H.M. Scheutjens, G.J. Fleer, Statistical thermodynamics of block copolymer adsorption. Part 2.—Effect of chain composition on the adsorbed amount and layer thickness, J. Chem. Soc., Farady Trans., 86 (1990) 1333-1340. doi: 1310.1039/ft9908601333.

[42] S. Banerjee, C. Cazeneuve, N. Baghdadli, S. Ringeissen, F.A.M. Leermakers, G.S. Luengo, Surfactant-polymer interactions: molecular architecture does matter, Soft Matter, 11 (2015) 2504-2511. doi: 2510.1039/c2505sm00117j.

[43] S. Banerjee, C. Cazeneuve, N. Baghdadli, S. Ringeissen, F. Lénforte, F.A.M. Leermakers, G.S. Luengo, Modeling of Polyelectrolyte Adsorption from Micellar Solutions onto Biomimetic Substrates, J. Phys. Chem. B, 121 (2017) 8638-8651. doi: 8610.1021/acs.jpcb.8637b05195.

[44] A. Naderi, P.M. Claesson, M. Bergström, A. Dédinaté, Trapped non-equilibrium states in aqueous solutions of oppositely charged polyelectrolytes and surfactants: effect of mixing protocol and salt concentration, Colloids Surf. A, 253 (2005) 83-93. doi: 10.1016/j.colsurfa.2014.1010.1123.

[45] I. Varga, R.A. Campbell, General Physical Description of the Behavior of Oppositely Charged Polyelectrolyte/Surfactant Mixtures at the Air/Water Interface, Langmuir, 33 (2017) 5915-5924. doi: 5910.1021/acs.langmuir.5917b01288.

[46] C.D. Bain, P.M. Claesson, D. Langevin, R. Meszaros, T. Nylander, C. Stubenrauch, S. Titmuss, R. von Klitzing, Complexes of surfactants with oppositely charged polymers at surfaces and in bulk, Adv. Colloid Interface Sci., 155 (2010) 32–49, doi: 10.1016/j.cis.2010.1001.1007.

[47] G. Nizri, S. Lagerge, A. Kamyshny, D.T. Major, S. Magdassi, Polymer–surfactant interactions: Binding mechanism of sodium dodecyl sulfate to poly(diallyldimethylammonium chloride), J. Colloid Interface Sci., 320 (2008 ) 74-81. doi:10.1016/j.jcis.2008.1001.1016.

[48] L. Braun, M. Uhlig, R. Von Klitzing, R.A. Campbell, Polymers and surfactants at fluid interfaces studied with specular neutron reflectometry, Adv. Colloid Interface Sci., 247 (2017) 130-148. doi: 110.1016/j.cis.2017.1007.1005.

[49] E. Staples, I. Tucker, J. Penfold, N. Warren, R.K. Thomas, D.J.F. Taylor, Organization of Polymer−Surfactant Mixtures at the Air−Water Interface: Sodium Dodecyl Sulfate and Poly(dimethyldiallylammonium chloride), Langmuir, 18 (2002) 5147-5153. doi: 5110.1021/la020034f.

[50] R.A. Campbell, A. Angus-Smyth, K. Yanez-Arteta, T. Tonigold, T. Nylander, I. Varga, New Perspective on the Cliff Edge Peak in the Surface Tension of Oppositely Charged Polyelectrolyte/Surfactant Mixtures, J. Phys. Chem. Lett., 1 (2010) 3021-3026. doi: 3010.1021/jz101179f.

[51] R.C. Oliver, J. Lipfert, D.A. Fox, R.H. Lo, S. Doniach, L. Columbus, Dependence of Micelle Size and Shape on Detergent Alkyl Chain Length and Head Group, Plos One, 8 (2013) e62488. doi: 62410.61371/journal.pone.0062488.

Adsorption of PDADMAC+RL mixtures

[52] N. Dhopatkar, J.H. Park, K. Chari, A. Dhinojwala, Adsorption and Viscoelastic Analysis of Polyelectrolyte–Surfactant Complexes on Charged Hydrophilic Surfaces, Langmuir, 31 (2015) 1026-1037. doi: 1010.1021/la5043052.

[53] R.A. Campbell, M.Y. Arteta, A. Angus-Smyth, T. Nylander, I. Varga, Multilayers at Interfaces of an Oppositely Charged Polyelectrolyte/Surfactant System Resulting from the Transport of Bulk Aggregates under Gravity, J. Phys. Chem. B, 116 (2012) 7981-7990. doi: 7910.1021/jp304564x.

[54] R.A. Campbell, M.Y. Arteta, A. Angus-Smyth, T. Nylander, B.A. Noskov, I. Varga, Direct Impact of Nonequilibrium Aggregates on the Structure and Morphology of Pdadmac/SDS Layers at the Air/Water Interface, Langmuir 30 (2014) 8664-8674. doi: 8610.1021/la500621t.

[55] F. Höök, J. Vörös, M. Rodahl, R. Kurrat, P. Böni, J.J. Ramsden, M. Textor, N.D. Spencer, P. Tengvall, J. Gold, B. Kasemo, A comparative study of protein adsorption on titanium oxide surfaces using in situ ellipsometry, optical waveguide lightmode spectroscopy, and quartz crystal microbalance/dissipation, Colloids Surf. B, 24 (2002) 155-170. doi: 110.1016/S0927-7765(1001)00236-00233.

[56] J. Vörös, The Density and Refractive Index of Adsorbing Protein Layers, Biophys. J., 87 (2004) 553–561. doi: 510.1529/biophysj.1103.030072.

[57] E. Guzmán, R. Chuliá-Jordán, F. Ortega, R.G. Rubio, Influence of the percentage of acetylation on the assembly of LbL multilayers of poly(acrylic acid) and chitosan, Phys. Chem. Chem. Phys., 13 (2011) 18200-18207. doi: 18210.11039/c18201cp21609k.

[58] M. Miyake, Y. Kakizawa, Morphological study of cationic polymer–anionic surfactant complex precipitated in solution during the dilution process, J. Cosmet. Sci., 61 (2010) 289-301.

[59] M. Miyake, Recent progress of the characterization of oppositely charged polymer/surfactant complex in dilution deposition system, Adv. Colloid Interface Sci., 239 (2017) 146-157. doi: 110.1016/j.cis.2016.1004.1007.

[60] N. Parimi, V. Viswanath, B. Kashyap, P.U. Patil, Hair as Biomarker of Fluoride Exposure in a Fluoride Endemic Area and a Low Fluoridated Area, Int. J. Trichology, 5 (2013) 148-150. doi: 110.4103/0974-7753.125613.

[61] N.A. Joshi, C.G. Ajithkrishnan, Scalp Hair as Biomarker for Chronic Fluoride Exposure among Fluoride Endemic and Low Fluoride Areas: A Comparative Study, Int. J. Trichology, 10 (2018) 71-75. doi: 10.4103/ijt.ijt_4191_4117.

*Supporting Information*

*for*

# Effect of molecular structure of eco-friendly glycolipid biosurfactants on the adsorption of hair-care conditioning polymers

Laura Fernández-Peña,[1] Eduardo Guzmán,[1,2,*] Fabien Leonforte,[3] Ana Serrano-Pueyo,[1] Krzysztof Regulski,[3] Lucie Tournier-Couturier,[3] Francisco Ortega,[1,2] Ramón G. Rubio[1,2,*] and Gustavo S. Luengo[3*]

[1]Departamento de Química Física I, Facultad de Ciencias Químicas

Universidad Complutense de Madrid, Ciudad Universitaria s/n, 28040-Madrid, Spain

[2] Instituto Pluridisciplinar, Universidad Complutense de Madrid

Paseo Juan XXIII, 1, 28040-Madrid, Spain

[3] L'Oréal Research and Innovation, Aulnay-Sous Bois, France

**S.1. Isolation and purification of Rhamnolipids**

**S.1.A. Obtention and purification of RL(C$_{10}$)**

*S.1.A.1. Microorganism maintenance and preinoculum preparation*

The PA-1 strain of *Pseudomonas aeruginosa* was preserved in 10 wt% glycerol in an ultrafreezer at -80°C following the procedure described by Santa Anna et al. [1]. The preinoculum was grown on a plate with YPDA (yeast extract 0.3 wt%, peptone 1.5 wt%, dextrose 0.1 wt%, agar 1.2 wt%) at 30°C for 48 hours and was transferred to 1000 ml flasks with 300 ml of a medium presenting the following composition (g/l): NaNO$_3$ 1, KH$_2$PO$_4$ 3, K$_2$HPO$_4$ 7, MgSO$_4$.7H$_2$O 0.2, yeast extract 5, peptone 5 and glycerol 30. After 24 hours of cultivation, the fermentation medium containing the cells was stored in cryotubes with a glycerol/fermentation medium ratio of 1:3 to act as standard preinoculum in all fermentations.

*S.1.A.2. Preinoculum preparation*

1.0 ml of the contents of a cryotube was inoculated into 300 ml of fermentation medium of the same composition to that used in the previous step. The flasks were then incubated on rotary agitators at 30°C and 170 rpm for 40 hours. At the end of this period the cells from each flask were recovered by centrifuging (5000 g for 20 minutes) and utilised as inoculum in the bioreactors.

*S.1.A.3. Sterilisation*

The fermenter, containing the culture medium to be utilised, was autoclave-sterilized at 121°C for 15 minutes prior to each production run. The oxygenation system was sterilised by circulating a 1.0 wt% solution of sodium hypochlorite for 1 hour. Following this procedure, sterile distilled water was circulated through the system to eliminate traces of chlorine. Only then the inoculation of the microorganisms was carried out.

*S.1.A.4. Production of rhamnolipids by Pseudomonas aeruginosa PA1*
Adsorption of PDADMAC+RL mixtures

The culture medium utilised in the fermentations had the following composition (g/l): glycerol 30, $NaNO_3$ 1, $K_2HPO_4$ 7, $KH_2PO_4$ 3 and $MgSO_4.7H_2O$ 0.2, there resulting a carbon/nitrogen ratio of 60. The principal fermentations were carried out in a BioFlo IIc bioreactor (Batch/Continuous Fermenter, New Brunswick Scientific, USA) of 5 L nominal capacity located in a fume cabinet with extraction. The average working volume utilised in the fermentations was 3 L. The temperature was maintained at $30^{\circ}C$ with agitation at 100 rpm. Oxygenation was carried out non-dispersively by means of a gas/liquid contactor, using compressed air or employing a cylinder of pure oxygen. Oxygenation conditions were initially defined in accordance with the results of the oxygenation tests with the module utilised.

*S.1.A.5. Sterilization and cell removal*

After the fermentation (7-9 days, in simple batches), the suspension was sterilized in an autoclave at 121°C for 15 min. The bacterial cells were removed by centrifugation, at 10,000 rpm for 20 minutes. After the cell removal, the water was removed by lyophilization.

*S.1.A.6. Rhamnolipid quantification and analytical data of the samples*

The rhamnolipid quantification was carried out in an indirect way, using rhamnose as a reference – rhamnose is a product of the acid hydrolysis of the rhamnolipids. For this purpose, a method adapted from that proposed by Pham et al. [2] was used.

The biosurfactant solution produced by this strain of *Pseudomonas aeruginosa* was also characterized by HPLC. A solution of $H_2SO_4$ was added to the biosurfactant solution until pH 2 was achieved, necessary for the precipitation of the rhamnolipids. The precipitated material was then dissolved in a chloroform:ethanol (2:1) solution and filtered. After the evaporation of the solvent, the purified rhamnolipids were derivatized and analyzed in the HPLC, using the patterns reported by Mata-Sandoval et al.[3]

**S.1.B. Protocol for obtaining RL($C_{14}$)**

Burkholderia thailandensis E264 was obtained from the American Type Culture Collection Adsorption of PDADMAC+RL mixtures

(ATCC® 700388TM). Bacteria were stored in nutrient broth (Becton Dickinson 234000) supplemented with 20 % glycerol at – 80 °C. For rhamnolipid production, cells from the stock were initially grown overnight in Tryptic Soy Broth (TSB) 6 g/L supplemented with 3 wt% glycerol at 30 °C and 160 rpm for pre-culture. Liquid cultures of 2 L Minimum Salt Medium (MSM: 4 g/L $Na_2HPO_4$, 1.5 g/L $KH_2PO_4$, 30 g/L glycerol 99%, 0.1 g/L citric acid, 0.005 g/L $FeSO_4$, 0.561 g/L urea, 0.05 g/L $MgSO_4$, pH 7) in 5 L flasks were inoculated with seed culture to reach an initial Optical Density ($OD_{600nm}$) equal to 0.1 and incubated at 30 °C for 10 days under gyratory shaking (110 rpm).

Extraction of total rhamnolipids was performed as follow: cells were removed from the medium by centrifugation (14500 g, 1h) and the supernatant was submitted to ultrafiltration (30 kDa, 0.1 m² polyethersulfone Sartorius). The retentate was then acidified to pH 2 using HCl 2N. The precipitate was successively washed with water and centrifuged to reach pH 4.5 and finally dried at 40°C under vacuum until steady weight.

Mono-RL($C_{14}$) and di-RL($C_{14}$) were purified from the total rhamnolipid extract using preparative column chromatography (Reveleris® PREP Purification system, Evaporative Light Scattering Detector).

## S.2. Preparation of PDADMAC-surfactant mixtures

PDADMAC-surfactant mixtures were prepared following a protocol adapted from that described by Llamas et al.[4], which can be summarized as follows: first the required amount of an aqueous stock solution of PDADMAC with concentration 20 wt% is weighted and poured into a flask to prepare a final mixture with a total polymer concentration of 0.2 wt%. Afterwards KCl (purity > 99.9 %) is added to reach a salt concentration in the final mixture of 40 mM. The Adsorption of PDADMAC+RL mixtures last step is the addition of the surfactant solution (pH ~ 5.6) and a subsequent dilution with acetic acid solution of pH ~ 5.6 up to reaching the desired surfactant concentration. It is worth mentioning that for the polyelectrolyte-surfactant solution preparation, the surfactant is added from solutions having a concentration one order of magnitude higher than that of the final mixture. The different components were successively added without any delay and the solutions were homogenized under mild stirring conditions (1000 rpm). It is expected that this procedure lead to kinetically-trapped aggregates during the initial mixing, as it has been commonly described for polyelectrolyte – surfactant mixtures.[5-8] In order to take into consideration its effect and to ensure reproducibility all samples were allowed aging during one week before their use at 25ºC.

## S.3. Experimental studies of the adsorption of polyelectrolyte-surfactant mixtures

### S.3.A. Characteristics of model surfaces

The model surfaces used in the present study are negatively charged, presenting a ξ-potential values around – 42 ± 5 mV (obtained from the measurement of the ξ-potential of colloidal particles with the same surface nature than the flat model surfaces [9]). Such values are in good agreement with those found for the ξ-potential of damaged hair fibres, from –55 to – 35 mV depending on their origin and chemical treatments.[10] Therefore, model surfaces can be used to mimic the hair fibre surface, at least from a physico-chemical perspective.

### S.3.B. Preparation of hair fibres for adsorption studies

Hair fibres were initially washed with a commercial shampoo, rinsed with tap water (35 ºC), and dried under ambient conditions. Then, fibres were bleached using a commercial bleaching product (persulfate based) during 30 minutes at room temperature. Afterwards, bleached fibres were rinsed abundantly with water and dried off.

Adsorption of PDADMAC+RL mixtures

For understanding the effect of the studied mixtures in hair fibres, images from hair fibres surface and transversal cut of them were obtained. For the later, fibres were previously stuffed in a Spurr low viscosity embedding resin (Sigma Aldrich, Germany) and then cut using a microtome [11].

## S.4. Foundations of SCF calculations

As input of the model, it is necessary to introduce the short-range interaction parameters (Flory-Huggins parameters χ) between segments of the molecules, dielectric permittivity (ε) and valence (ν). The valence accounts for the electrostatic contributions, whereas the Flory-Huggins parameters play a central role in the formation and stability of the self-assembly colloids. All these parameters are given in Table S.1. It is also worth to mention that a five, 5-sites description of water molecules in the used scheme has been chosen in order to mimic water hydrogen-bonding ability, especially with O and OH groups of sugar rings, and with negative Flory-Huggins parameters that ensure solubility of species. For the sulfonate groups of SLES, a 5 sites representation has been used, as well as for cationic groups of PDADMAC, i.e. N, and deprotonated segment of carboxylic groups in rhamnolipids.

Furthermore, for SCF calculations, a discretization of the molecules is needed. For this purpose a lattice size $a = 0.3$ nm was chosen in agreement with the discussion by Banerjee et al.[12, 13] Furthermore, the values for the Flory-Huggins parameter used in the current work were adapted from ref. [12, 13]. The mean-field free-energy functional can be written in terms of molecule's segment density profiles for a specific segment type, $\varphi_X(z)$, and conjugate segment potential profiles, $u_X(z)$, with $X$ being referred to a segment type and $z$ indicating the spatial coordinate. The optimization process is based in a numerical SCF procedure which allows us to relate the volume fractions and the potentials following the basic principles described in the literature.[12-14]

Table S.1. Flory-Huggins Interaction parameters, $\chi$, between various pairs of segments, relative dielectric constant, $\varepsilon$, and valency, $v$, of the segment types as used in the SCF calculations. Note that the segment types X denote the deprotonated segment of carboxylic group, and Si denotes the surface.

| $\chi$ | $w$ | $C$ | $O$ | $S$ | $N$ | $K$ | $Cl$ | $OH$ | $X$ | $Si$ | $\varepsilon$ | $v$ |
|---|---|---|---|---|---|---|---|---|---|---|---|---|
| $w$ | 0 | 1.6 | -0.6 | 0 | 0.5 | 0 | 0 | -0.6 | 0 | 1 | 80 | 0 |
| $C$ | 1.6 | 0 | 1.6 | 2 | 2 | 2 | 2 | 2 | 2 | 0 | 2 | 0 |
| $O$ | -0.6 | 1.6 | 0 | 0 | 0 | 0 | 0 | 0 | 0 | 0 | 1.5 | 0 |
| $S$ | 0 | 2 | 0 | 0 | 0 | 0 | 0 | 0 | 0 | 0 | 3.4 | -0.2 |
| $N$ | 0.5 | 2 | 0 | 0 | 0 | 0 | 0 | 0 | 0 | 0 | 7 | 0.2 |
| $K$ | 0 | 2 | 0 | 0 | 0 | 0 | 0 | 0 | 0 | 0 | 6.1 | 1 |
| $Cl$ | 0 | 2 | 0 | 0 | 0 | 0 | 0 | 0 | 0 | 0 | 6.1 | -1 |
| $OH$ | -0.6 | 2 | 0 | 0 | 0 | 0 | 0 | 0 | 0 | 0 | 1.5 | 0 |
| $X$ | 0 | 2 | 0 | 0 | 0 | 0 | 0 | 0 | 0 | 0 | 1.5 | -0.2 |
| $Si$ | 1 | 0 | 0 | 0 | 0 | 0 | 0 | 0 | 0 | 0 | 5 | -0.1 |

The molecular partition functions for single chains play a central role in the mean field free-energy. From the operational point of view, it is convenient to use the dimensionless form of the Edward diffusion equation (1) [15]

$$\frac{\partial G(z,s)}{\partial s} = \frac{1}{6}\frac{\partial G^2(z,s)}{\partial z^2} - u(z,s)G(z,s), \qquad (S.1)$$

Adsorption of PDADMAC+RL mixtures

where $s$ and $u(z,s)$ are referred to the number of segments and to the volume fraction for a given segment potential, respectively. $G(z,s)$ represents the end-point distribution for a chain segment which provides information related to the statistical weight that the end point of a walk formed big $s$ segments ends up at position $z$. $G(z,s)$ can be assumed to be the Boltzmann weight of the field.[16] Equation (S.1) is solved considering the existence of a hydrophilic surface for $z = 0$ and that for large $z$ values the behavior of the mixture is similar to that found in the bulk as boundary conditions. The SCF approach requires the use of freely jointed chains (FJC) to map the Edward equation over a lattice, i.e. considers freely rotating bonds with the same length [17]. The use of this FJC model is preferred since it allows one to obtain the partition function and the volume fraction for a given segment potential $u(z,s)$. Further details are described in Ref. [13].

The segment potentials can be computed when the volume fractions are available. The Flory-Huggins-like interaction parameter and a contribution to ensure the compressibility of the system are used to calculate the interaction energy of a segment at a specified location. The estimation of the number of segment-segment contacts is calculated using the Bragg-Williams mean-field approximation.[18] In addition, in the segment potential the appearance of terms accounting for the long-range electrostatic interactions must be included, which makes it necessary to solve the Poisson equation

$$\frac{\partial \varepsilon(z)}{\partial z} \frac{\partial \psi(z)}{\partial z} = -q(z) \qquad (S.2)$$

with $\varepsilon(z)$ and $\dfrac{\partial \psi(z)}{\partial z}$ being the dependence of the dielectric constant on the position and the electrostatic position $\psi(z)$ gradient, respectively. $q(z)$ defines the charge density $q(z) = \sum_X \varphi_X e v_X$ , where $e$ is the elementary charge and $V_X$ the valence including the sign of the

segment $X$. Before performing the analysis of the adsorption onto the surface, the bulk

composition and aggregation were optimized based on SCF calculations according to Banerjee

et al.[12]. This allows us to obtain the volume fraction profiles for a segment number $s$ at

coordinate $z$, $\varphi_i(z)$. The adsorption is quantified in terms of the excess adsorption

$$\theta_i^\sigma = \sum_{z=1}^M \left[ \varphi_i(z) - \varphi_i^b \right] ,$$ where $\varphi_i^b$ is the volume fraction in the bulk obtained from SCF

optimization of the bulk [12].

For the SCF calculations, a configuration considering the formation of polyelectrolyte-surfactant spherical micelles, i.e. the formation of polyelectrolyte-surfactant complexes in which spherical micelles of a surfactant form a core surrounded by the polyelectrolyte chains, were chose [12]. For mixtures, the grand potential $\Omega$, which informs about the surfactant concentration at fixed polymer concentration, was kept fixed to $5k_BT$, i.e. above the value at which the first micelle appears. The choice of the grand potential per micelle does not modify significantly the conclusions, and only modifies slightly the number of micelles in solution.

## S.5. Results

### S.5.A. _Experimental studies on the association of PDADMAC-surfactant in solution_

Adsorption of PDADMAC+RL mixtures

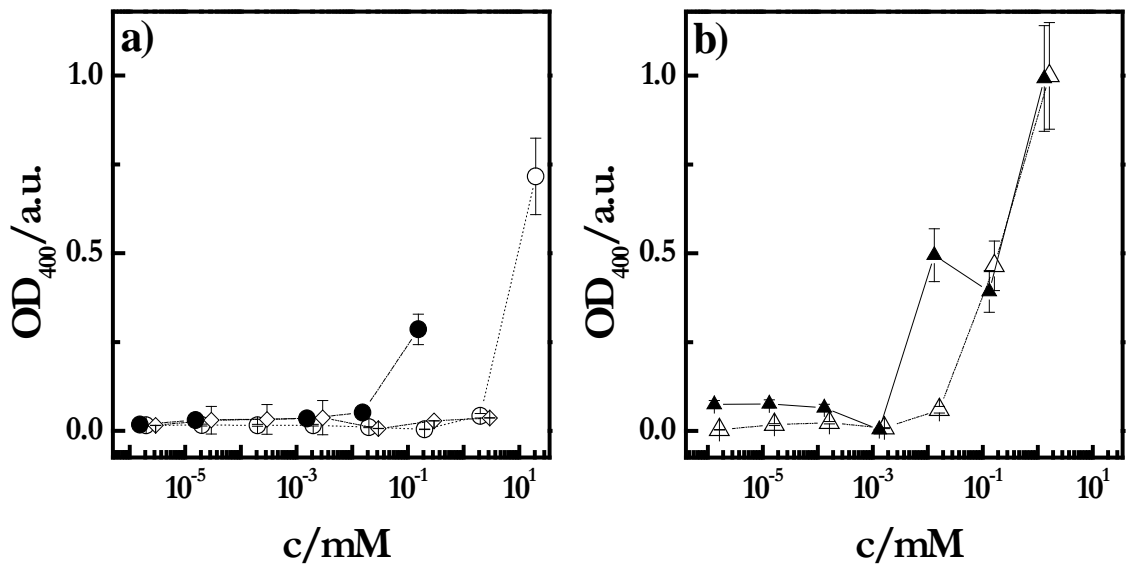

Figure S.1. Dependences of the optical density measured at 400 nm on c for different PDADMAC – surfactant mixtures: (a) Effect of the hydrophilicity and charge for $C_{10}$ surfactants. (b) Effect of the hydrophilicity for $C_{14}$ surfactants. The symbols are referred to different samples: ○ mono-RL($C_{10}$), Δ mono-RL($C_{14}$), ● di-RL($C_{10}$), ▲ di-RL($C_{14}$) and ◊APG. The lines are guides for the eyes. The results correspond to PDADMAC-surfactant mixtures containing a fixed PDADMAC concentration of 0.5 wt% (pH = 5.6 and 40 mM of KCl concentration), and left to age for one week prior to measurement.

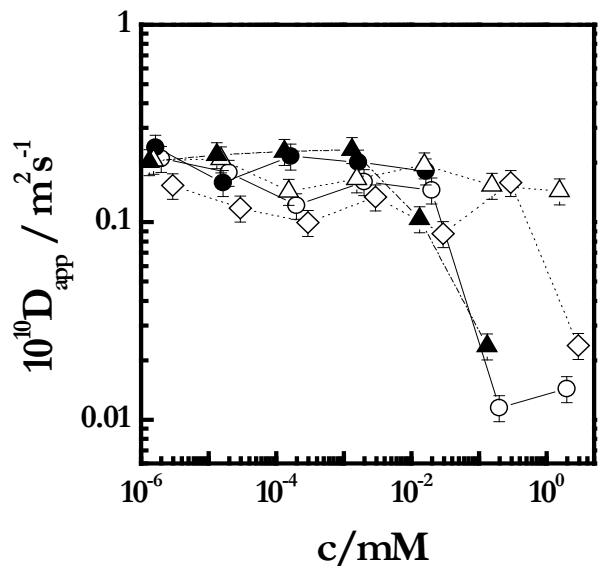

Figure S.2. Surfactant concentration dependence of $D_{app}$ for the different mixtures studied. The symbols are referred to different samples: ○ mono-RL($C_{10}$), △ mono-RL($C_{14}$), ● di-RL($C_{10}$), ▲ di-RL($C_{14}$) and ◊APG. The lines are guides for the eyes. The results correspond to PDADMAC-surfactant mixtures containing a fixed PDADMAC concentration of 0.5 wt% (pH = 5.6 and 40 mM of KCl concentration), and left to age for one week prior to measurement.

*S.4.B. SCF calculations on the association of PDADMAC-surfactant in solution*

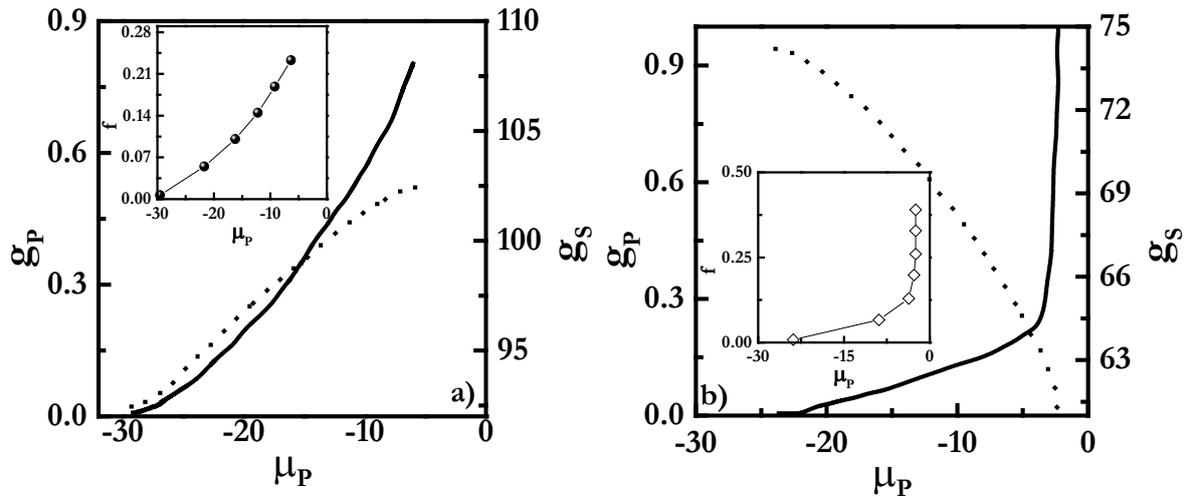

Figure S.3. Binding isotherms for PDADMAC-SLES (a) and PDADMAC-APG (b) mixturesrepresented as the dependence on the PDADMAC chemical potential, $\mu_P$, of the aggregation number of the surfactant in the complexes, $g_S$, (dotted lines) and the degree of polyelectrolyte-surfactant binding, $g_P$, (solid lines). The inserted panels represent the dependence of the composition ratio $f$ in the complexes on $\mu_P$ for the same mixtures shown in the main panel. Note that $\mu_P$ is equivalent to the amount of polymer in the complexes and that the calculations were performed under spherical micelles configurations and fixed free energy $5k_BT$.

Adsorption of PDADMAC+RL mixtures

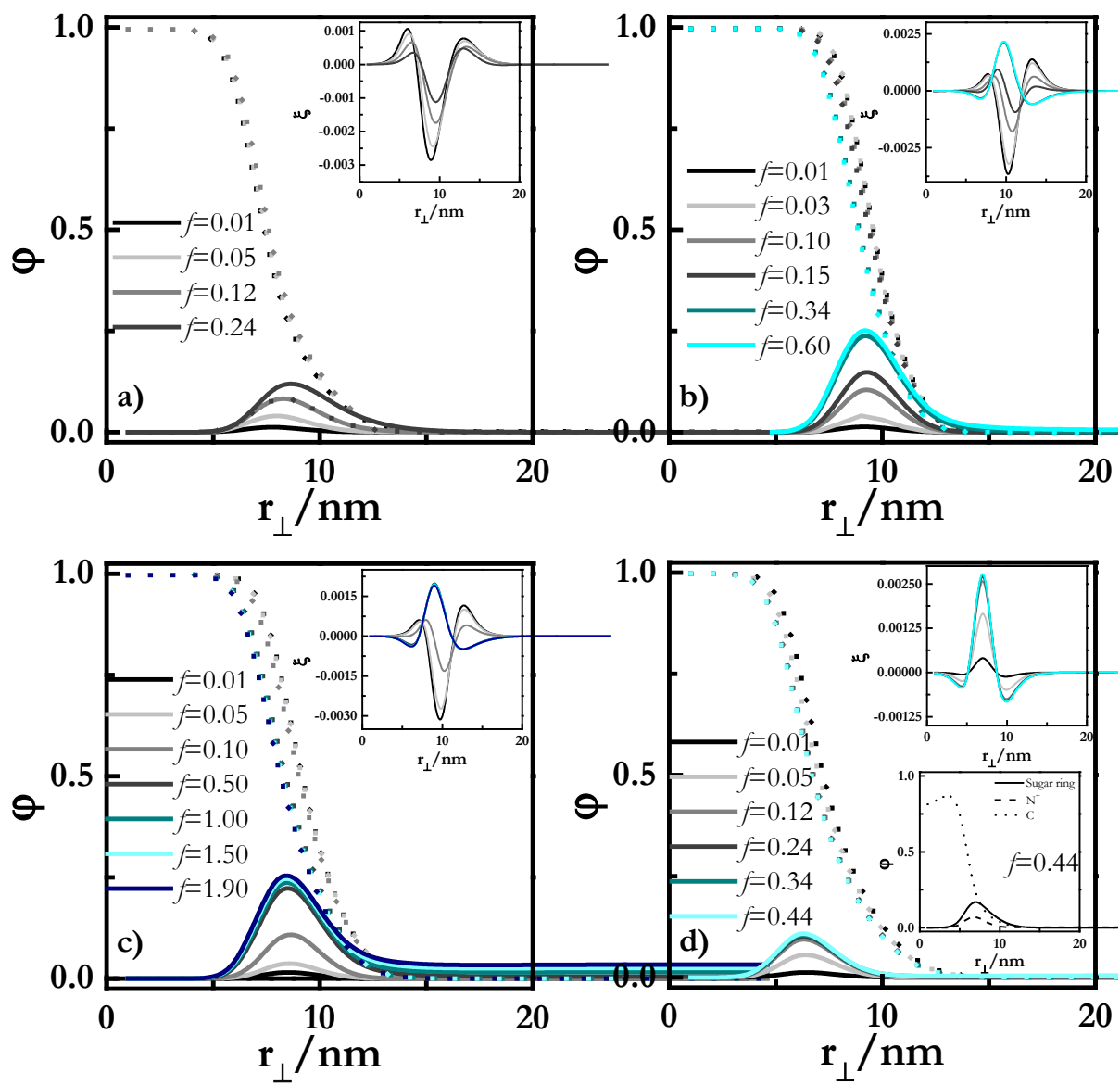

Figure S.4. Radial volume fraction profiles for PDADMAC-surfactant mixtures: PDADMAC-SLES (a), PDADMAC- mono-RL (b, both $C_{10}$ and $C_{14}$ are shown) and PDADMAC- di-RL (c, both $C_{10}$ and $C_{14}$ are shown) and PDADMAC-APG (d) as a function of the composition fraction, $f$. The insets represent the charge density profiles for each mixture as a function of $f$. In panel (b), an additional inset showing the radial volume fraction profiles of sugar ring, ammonium and hydrophobic groups in the complexes for a fixed value of $f = 0.44$.

*S.4.C. Experimental studies on the adsorption of PDADMAC-surfactant onto solid surfaces*

Adsorption of PDADMAC+RL mixtures

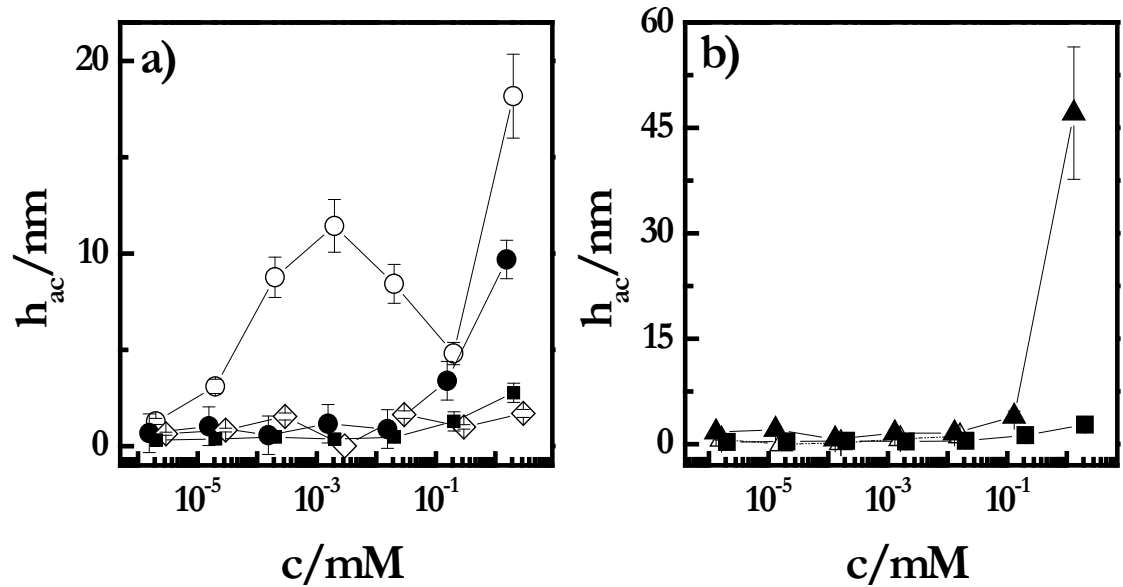

Figure S.5 $h_{ac}$ dependences on c for the different PDADMAC–surfactant mixtures studied: (a) Effect of the hydrophilicity and charge for $C_{10}$ surfactants and (b) Effect of the hydrophilicity for $C_{14}$ surfactants. The symbols are referred to different samples: ○ mono-RL($C_{10}$), Δ mono-RL($C_{14}$), ● di-RL($C_{10}$), ▲ di-RL($C_{14}$) and ◊APG. In all the plots, the data corresponding to samples PDADMAC – SLES are included (■).The lines are guides for the eyes. The results correspond to PDADMAC-surfactant mixtures containing a fixed PDADMAC concentration of 0.5 wt% (pH = 5.6 and 40 mM of KCl concentration), and left to age for one week prior to measurement.

Adsorption of PDADMAC+RL mixtures

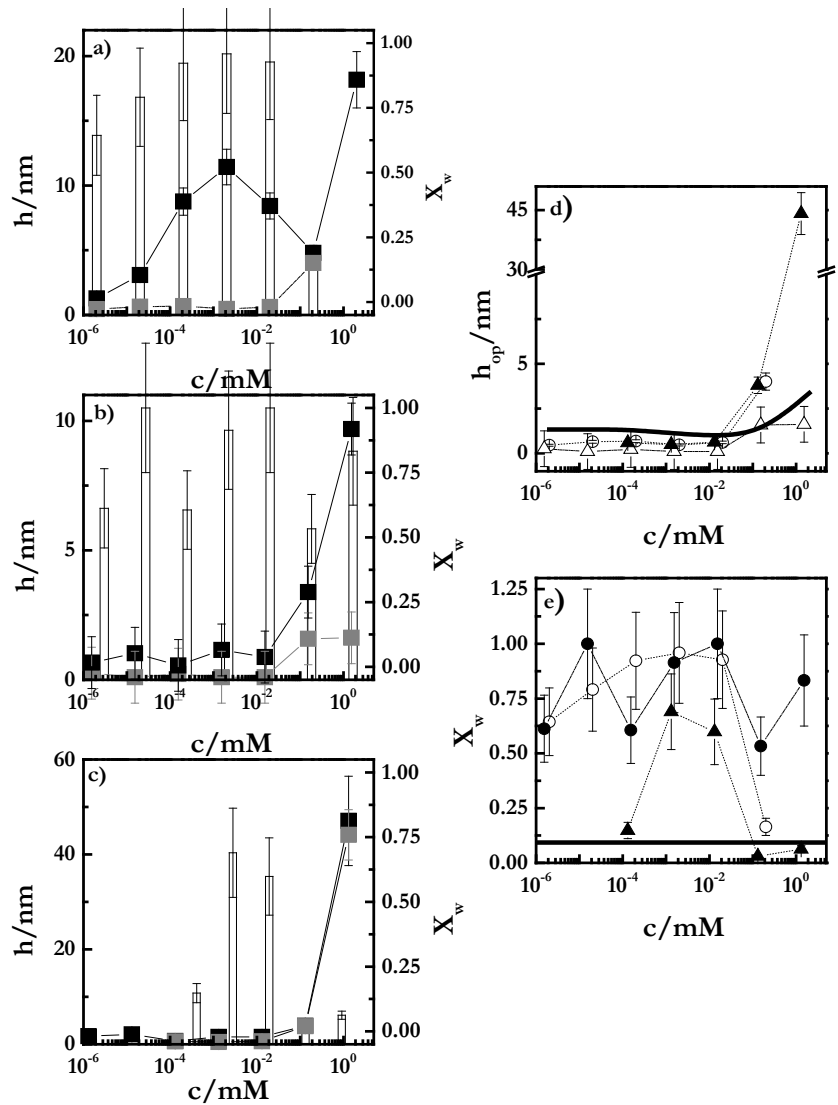

Figure S.6. (a)-(c) $h_{ac}$ (■) and $h_{op}$ dependences (▩) on c for PDADMAC – surfactant mixtures studies (left axis) and water content, $X_w$, dependences on c (white bars, right axis) for (a) PDADMAC - mono-RL($C_{10}$) (a), PDADMAC - di-RL($C_{10}$) (b) and PDADMAC - di-RL($C_{14}$) (c). (d) Concentration dependences of $h_{op}$ for binary mixtures of PDADMAC and SLES (solid line), and mono-RL($C_{10}$) (○), di-RL($C_{10}$) (●) and di-RL($C_{14}$) (▲). (e) Concentration dependences of $X_w$ for binary mixtures of PDADMAC and SLES (solid line), and mono-RL($C_{10}$) (○), di-RL($C_{10}$) (●) and di-RL($C_{14}$) (▲). The results correspond to PDADMAC-surfactant mixtures containing a fixed PDADMAC concentration of 0.5 wt% (pH = 5.6 and 40 mM of KCl concentration), and left to age for one week prior to measurement.

Adsorption of PDADMAC+RL mixtures

*S.4.D. <u>SCF calculations on the adsorption of PDADMAC-surfactant onto solid surfaces</u>*

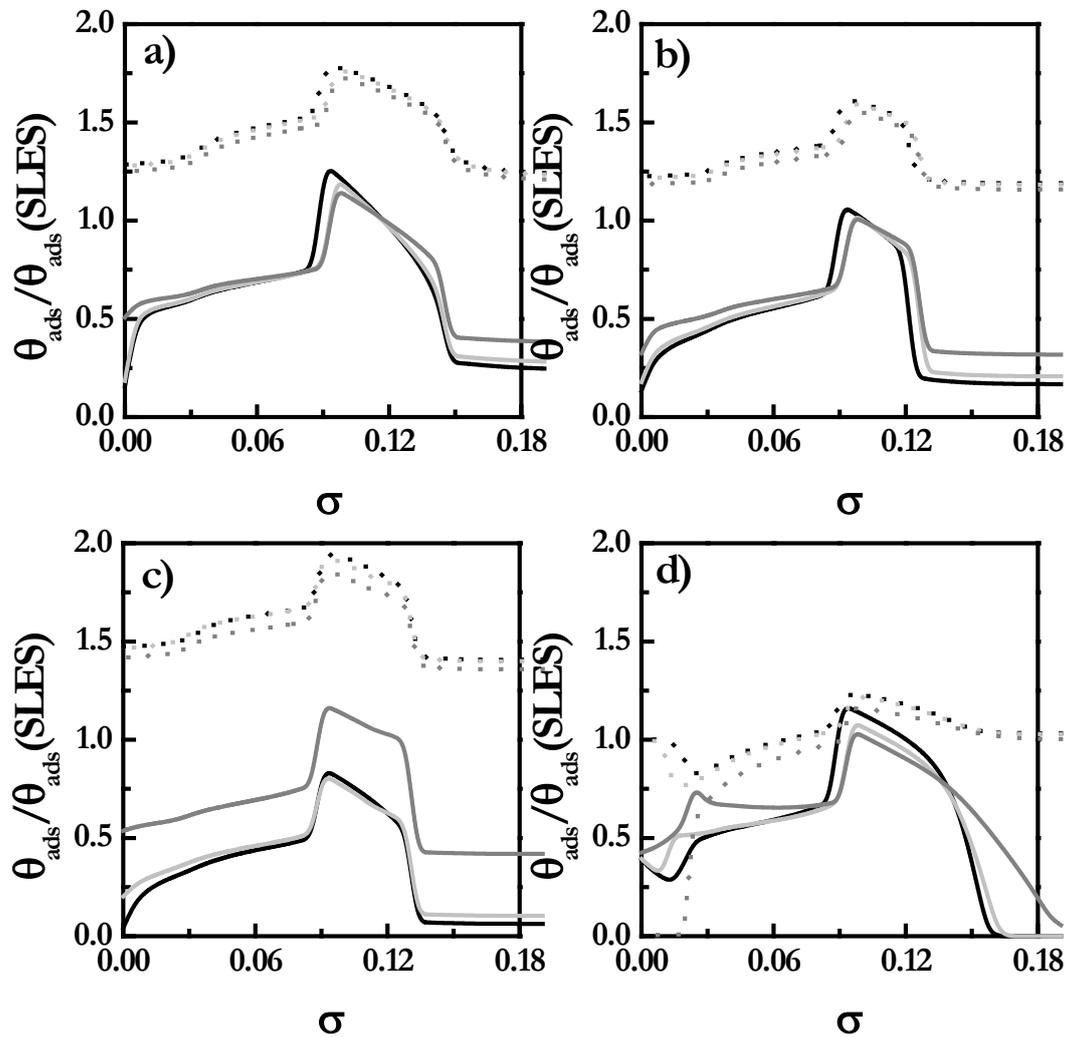

Figure S.7. Dependence of the ratio of the adsorbed amount of surfactants (dotted lines) and PDADMAC (solid lines) to SLES adsorbed amount on the hydrophobicity degree of the surface $\sigma$ for the same values of $f$ (— and ···· $f = 5\cdot10^4$, — and ···· $f = 5\cdot10^3$, — and ···· $f = 5\cdot10^2$): PDADMAC - mono-RL(C$_{10}$) (a), PDADMAC – di-RL(C$_{10}$) (b), PDADMAC – di - RL(C14) (c) and PDADMAC-APG (d),. Different values of the composition fractions, $f$, are considered, and calculations were done from purely hydrophilic charged surface ($\sigma = 0$) to increasingly hydrophobic charged surface ($\sigma > 0$).

Adsorption of PDADMAC+RL mixtures

*S.4.E. Adsorption of polymer-surfactant mixtures onto hair fibers*

The SEM image of the uncoated fiber (Figure S.8a) shows the presence of the characteristic scales of the hair cuticle, which presents partial delamination and breakage. Once the fiber is treated during 15 minutes with the polyelectrolyte - rhamnolipid mixture, partial coating of the scales was found (see Figure S.8b). This is expected to present a key role in the improvement of the manageability and reduction of the friction between fibers. The efficiency of the coating with the polyelectrolyte-surfactant mixture is also evidenced from the EDS spectra. The analysis of the EDS spectrum of the untreated fiber shows the appearance of C, N, O and S typical of hair keratin was found in the analysis, whereas the absence of S was found in the spectrum of treated fiber. This may be explained considering that the adsorption of the polyelectrolyte - surfactant mixture onto the fiber surface leads to the formation of a layer thicker than the penetration depth of the secondary X-ray and consequently the S of the hair surface is not found in the spectra. It is worth mentioning that in both cases gold was found in the EDS analysis, which is explained considering that it is necessary coating the samples with a conductive gold layer previously to the SEM observation. The results obtained by SEM confirm the high efficiency of the adsorption of polyelectrolyte-surfactant mixtures onto the hair fiber. The comparison of Figure S.8b and S.8c point out that the increase of the adsorption time leads to a significant enhancement of the deposition, with the scale of the hair cuticle being completely coated.

Adsorption of PDADMAC+RL mixtures

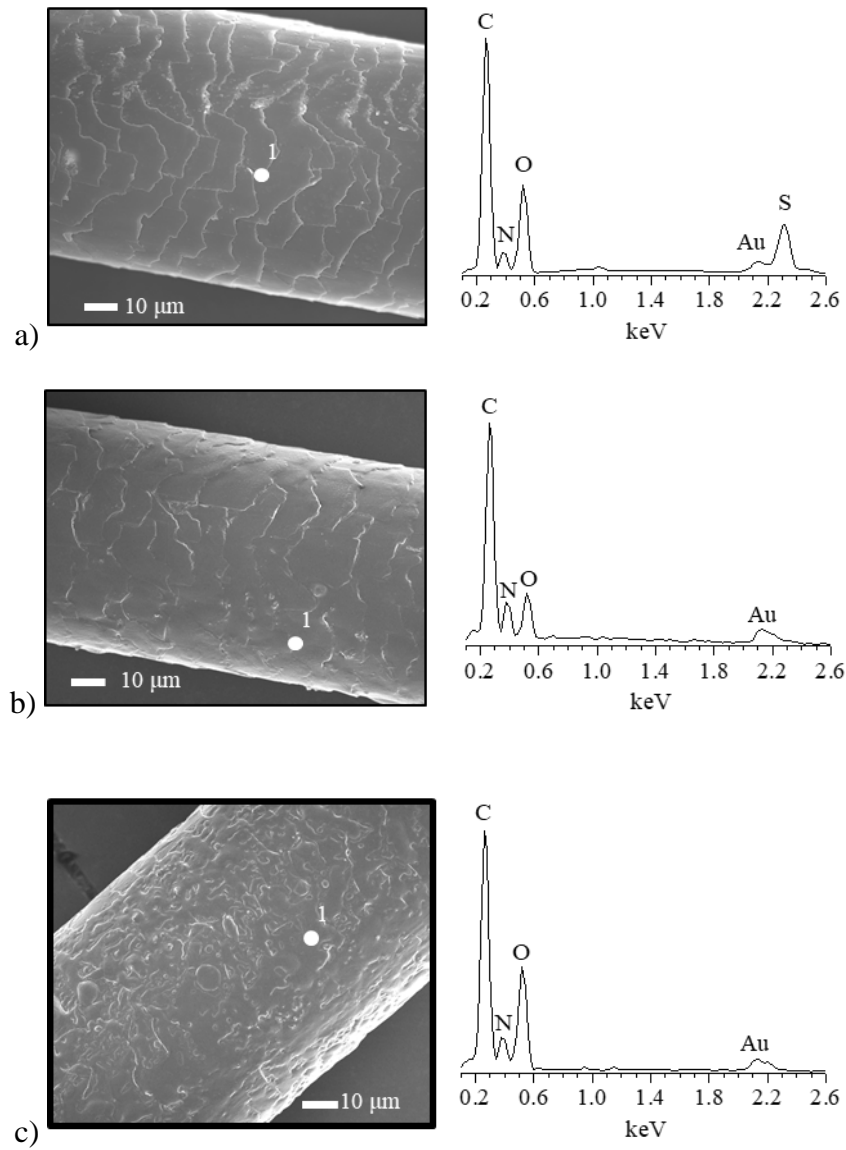

Figure S.8. SEM images and their respective EDS spectra for hair fibers uncoated (a) and coated (b, treatment during 15 minutes) and (c, treatment during 2 hours) with a polyelectrolyte - rhamnolipid mixture with a surfactant concentration of 1 mM. The results correspond to PDADMAC-surfactant mixtures containing a fixed PDADMAC concentration of 0.5 wt% (pH = 5.6 and 40 mM of KCl concentration), and left to age for one week prior to measurement.

Adsorption of PDADMAC+RL mixtures

# References

[1] L.M. Santa Anna, G.V. Sebastian, E.P. Menezes, T.L.M. Alves, A.S. Santos, N. Pereira Jr., D.M.G. Freire, Production of biosurfactants from Pseudomonas aeruginosa PA1 isolated in oil environment, Braz. J. Chem. Eng., 19 (2002) 159-166. doi: 110.1590/S0104-66322002000200011.

[2] T.H. Pham, J.S. Webb, B.H.A. Rehm, The role of polyhydroxyalkanoate biosynthesis by Pseudomonas aeruginosa in rhamnolipid and alginate production as well as stress toleranceand biofilm formation, Microbiology 150 (2004) 3405-34213. doi: 34210.31099/mic.34210.27357-34210.

[3] J.C. Mata-Sandoval, J. Karns, A. Torrens, High-performance liquid chromatography method for the characterization of rhamnolipid mixtures produced by pseudomonas aeruginosa UG2 on corn oil., J. Cromatogr. A 864 (1999) 211-220. doi: 210.1016/S0021-9673(1099)00979-00976.

[4] S. Llamas, E. Guzmán, N. Baghdadli, F. Ortega, C. Cazeneuve, R.G. Rubio, G.S. Luengo, Adsorption of poly(diallyldimethylammonium chloride)—sodium methyl-cocoyl-taurate complexes onto solid surfaces, Colloids Surf. A, 505 (2016) 150-157. doi: 110.1016/j.colsurfa.2016.1003.1003.

[5] A. Angus-Smyth, C.D. Bain, I. Vargac, R.A. Campbell, Effects of bulk aggregation on PEI–SDS monolayers at the dynamic air–liquid interface: depletion due to precipitation versus enrichment by a convection/spreading mechanism, Soft Matter, 9 (2013) 6103-6117. doi: 6110.1039/c6103sm50636c.

[6] R.A. Campbell, M. Yanez-Arteta, A. Angus-Smyth, T. Nylander, B.A. Noskov, I. Varga, Direct Impact of Non-Equilibrium Aggregates on the Structure and Morphology of Pdadmac/SDS Layers at the Air/Water Interface, Langmuir, 30 (2014) 8664–8674. doi: 8610.1021/la500621t.

[7] R. Mészáros, L. Thompson, M. Bos, I. Varga, T. Gilányi, Interaction of sodium dodecyl sulfate with polyethyleneimine: surfactant-induced polymer solution colloid dispersion transition, Langmuir 19 (2003) 609-615. doi: 610.1021/la026616e.

[8] A. Mezei, R. Meszaros, I. Varga, T. Gilanyi, Effect of Mixing on the Formation of Complexes of Hyperbranched Cationic Polyelectrolytes and Anionic Surfactants, Langmuir, 23 (2007) 4237-4247. doi: 4210.1021/la0635294

[9] E. Guzmán, H. Ritacco, J.E.F. Rubio, R.G. Rubio, F. Ortega, Salt-induced changes in the growth of polyelectrolyte layers of poly(diallyldimethylammoniumchloride) and poly(4-styrene sulfonate of sodium), Soft Matter, 5 (2009) 2130-2142. doi: 2110.1039/b901193e.

[10] G.S. Luengo, A. Galliano, C. Dubief, Aqueous Lubrication in Cosmetic, in: N.D. Spencer (Ed.) Aqueous Lubrication. Natural and Biomimetic Approaches, World Scientific Publishing Co. Pte. Ltd., Singapore, Singapore, 2014, pp. pp 103-144. doi: 110.1142/9789814313773_9789814310004.

[11] W.M. Hess, R.E. Seegmiller, Computerized Image Analysis of Resin-Embedded Hair Trans. Am. Microsc. Soc. , 107 (1988) 421-425. doi: 410.2307/3226338.

[12] S. Banerjee, C. Cazeneuve, N. Baghdadli, S. Ringeissen, F.A.M. Leermakers, G.S. Luengo, Surfactant-polymer interactions: molecular architecture does matter, Soft Matter, 11 (2015) 2504-2511. doi: 2510.1039/c2505sm00117j.

[13] S. Banerjee, C. Cazeneuve, N. Baghdadli, S. Ringeissen, F. Lénforte, F.A.M. Leermakers, G.S. Luengo, Modeling of Polyelectrolyte Adsorption from Micellar Solutions onto Biomimetic Substrates, J. Phys. Chem. B, 121 (2017) 8638-8651. doi: 8610.1021/acs.jpcb.8637b05195.

Adsorption of PDADMAC+RL mixtures

[14] G. Fleer, M.A. Cohen-Stuart, J. Scheutjens, T. Crosgrove, B. Vincent, Polymer at Interfaces, Chapman and Hall, London, United Kingdom, 1993.

[15] S. Edwards, The Statistical Mechanics of Polymers with Excluded Volume, Proc. Phys. Soc., London, 85 (1965) 613-624. doi: 610.1088/0370-1328/1085/1084/1301.

[16] G.J. Fleer, Polymers at interfaces and in colloidal dispersions, Adv. Colloid Interface Sci., 159 (2010 ) 99-116. doi: 110.1016/j.cis.2010.1004.1004.

[17] M. Rubinstein, R.H. Colby, Polymer Physics., Oxford University Press New York, United States of America, 2003.

[18] T. Hill, An Introduction in Statistical Thermodynamics, Wesley Publisher, London, United Kingdom, 1960.